\documentclass[twocolumn]{aastex62}

\newcommand{\kms}{km~s$^{-1}$}

\slugcomment{To appear in The Astrophysical Journal}

\shorttitle{Ultra-Diffuse Galaxies in the Virgo Cluster}
\shortauthors{Lim et al.}

\begin{document}


\title{\bf \large The Next Generation Virgo Cluster Survey (NGVS). XXX. Ultra-Diffuse Galaxies and their Globular Cluster Systems}


\correspondingauthor{Sungsoon Lim}
\email{sslim00@gmail.com}

\author[0000-0002-5049-4390]{Sungsoon Lim}
\affiliation{Herzberg Astronomy and Astrophysics Research Centre, National Research Council of Canada, Victoria, BC V9E 2E7, Canada}
\affiliation{University of Tampa, 401 West Kennedy Boulevard, Tampa, FL 33606, USA}

\author[0000-0003-1184-8114]{Patrick C{\^o}t{\'e}}
\affiliation{Herzberg Astronomy and Astrophysics Research Centre, National Research Council of Canada, Victoria, BC V9E 2E7, Canada}

\author[0000-0002-2073-2781]{Eric W. Peng}
\affiliation{Department of Astronomy, Peking University, Beijing 100871, China}
\affiliation{Kavli Institute for Astronomy and Astrophysics, Peking University, Beijing 100871, China}

\author[0000-0002-8224-1128]{Laura Ferrarese}
\affiliation{Herzberg Astronomy and Astrophysics Research Centre, National Research Council of Canada, Victoria, BC V9E 2E7, Canada}

\author[0000-0002-0363-4266]{Joel C. Roediger}
\affiliation{Herzberg Astronomy and Astrophysics Research Centre, National Research Council of Canada, Victoria, BC V9E 2E7, Canada}

\author[0000-0001-9427-3373]{Patrick R. Durrell}
\affiliation{Department of Physics and Astronomy, Youngstown State University, One University Plaza, Youngstown, OH 44555, USA}

\author[0000-0002-7089-8616]{J. Christopher Mihos}
\affiliation{Department of Astronomy, Case Western Reserve University, 10900 Euclid Ave, Cleveland, OH 44106, USA}

\author[0000-0003-4672-8497]{Kaixiang Wang}
\affiliation{Department of Astronomy, Peking University, Beijing 100871, China}
\affiliation{Kavli Institute for Astronomy and Astrophysics, Peking University, Beijing 100871, China}

\author{S.D.J. Gwyn}
\affiliation{Herzberg Astronomy and Astrophysics Research Centre, National Research Council of Canada, Victoria, BC V9E 2E7, Canada}

\author{Jean-Charles Cuillandre}
\affiliation{AIM Paris Saclay, CNRS/INSU, CEA/Irfu, Université Paris Diderot, Orme des Merisiers, F-91191 Gif-sur-Yvette Cedex, France}

\author[0000-0002-4718-3428]{Chengze Liu}
\affiliation{Department of Astronomy, School of Physics and Astronomy, and Shanghai Key Laboratory for Particle Physics and Cosmology, Shanghai Jiao Tong University, Shanghai 200240, China}

\author[0000-0003-4945-0056]{Rub{\'e}n S{\'a}nchez-Janssen}
\affiliation{STFC UK Astronomy Technology Centre, Royal Observatory, Blackford Hill, Edinburgh, EH9 3HJ, UK}

\author[0000-0001-6443-5570]{Elisa Toloba}
\affiliation{Department of Physics, University of the Pacific, 3601 Pacific Avenue, Stockton, CA 95211, USA}

\author[0000-0002-3790-720X]{Laura V. Sales}
\affiliation{Department of Physics and Astronomy, 900 University Avenue, Riverside, CA 92521, USA)}

\author[0000-0001-8867-4234]{Puragra Guhathakurta}
\affiliation{UCO/Lick Observatory, University of California, Santa Cruz, 1156 High Street, Santa Cruz, CA 95064, USA}

\author{Ariane Lan{\c c}on}
\affiliation{Universit{\'e} de Strasbourg, CNRS, Observatoire Astronomique de Strasbourg, UMR 7550, 11 rue de l'Universit{\'e}, F-67000 Strasbourg, France}

\author[0000-0003-0350-7061]{Thomas H. Puzia}
\affiliation{Institute of Astrophysics, Pontificia Universidad Cat{\'o}lica de Chile, Av. Vicu{\~n}a Mackenna 4860, 7820436 Macul, Santiago, Chile}



\begin{abstract}
We present a study of ultra-diffuse galaxies (UDGs) in the Virgo Cluster based on deep imaging from the Next Generation Virgo Cluster Survey (NGVS). Applying a new definition for the UDG class based on galaxy scaling relations, we define samples of 44 and 26 UDGs using expansive and restrictive selection criteria, respectively. Our UDG sample includes objects that are significantly fainter than previously known UDGs: i.e., more than half are fainter than $\langle\mu\rangle_e \sim27.5$ mag arcsec$^{-2}$. The UDGs in Virgo's core region show some evidence for being structurally distinct from ``normal" dwarf galaxies, but this separation disappears when considering the full sample of galaxies throughout the cluster. UDGs are more centrally concentrated in their spatial distribution than other Virgo galaxies of similar luminosity, while their morphologies demonstrate that at least some UDGs owe their diffuse nature to physical processes---such as tidal interactions or low-mass mergers---that are at play within the cluster environment. The globular cluster (GC) systems of Virgo UDGs have a wide range in specific frequency ($S_N$), with a higher mean $S_N$ than ``normal'' Virgo dwarfs, but a lower mean $S_N$ than Coma UDGs at fixed luminosity. Their GCs are predominantly blue, with a small contribution from red clusters in the more massive UDGs. The combined GC luminosity function is consistent with those observed in dwarf galaxies, showing no evidence of being anomalously luminous. The diversity in their morphologies and their GC properties suggests no single process has given rise to all objects within the UDG class. Based on the available evidence, we conclude that UDGs are simply those systems that occupy the extended tails of the galaxy size and surface brightness distributions. 
\end{abstract}


\keywords{galaxies: clusters: individual (Virgo) --- galaxies: formation --- galaxies: evolution --- galaxies: star clusters: general}



\section{Introduction}

Ultra-diffuse galaxies (UDGs) are a class of low surface brightness (LSB) galaxies that have luminosities characteristic of dwarf galaxies, but sizes more typical of giants. Although the existence of such galaxies was established in the 1980s \citep{Rea83,Bin85,Imp88}, interest in them surged following the discovery of 47 faint, diffuse galaxies in the Coma Cluster by \citet{vanD15}, who introduced the UDG terminology. Many UDG candidates have now been discovered in both cluster (e.g., \citealp{Mih15,Mun15,vanB16,Lee17,Mih17,Jan17,Ven17,Wit17}), and group or field environments (e.g., \citealp{Mar16,Mer16,Rom17,Tru17,Gre18,Pro19a}).

The very existence of UDGs in high-density regions, such as rich clusters, prompts the question: how can such faint and diffuse galaxies survive in these hostile environments? While the physical properties and overall numbers of these galaxies remain uncertain, several scenarios have been proposed to explain their origin, with no clear consensus having yet emerged. On one hand, estimates of their total (gravitating) mass based on their globular cluster (GC) systems seem to suggest that at least some UDGs are embedded in massive dark matter halos ($10^{11} M_{\odot} \lesssim M_{DM} \lesssim 10^{12} M_{\odot}$; e.g., \citealp{Bea16,Pen16,vanD17,Tol18}) that allow them to survive in dynamically hostile environments. If this inference is correct, then one could consider UDGs to be ``failed'' galaxies \citep{Yoz15}: i.e., dark matter dominated systems that were, for some reason, inefficient in assembling (or retaining) the stellar components typical of most galaxies with such massive dark matter halos. On the other hand, the discovery of some UDGs in low-density environments suggests that such systems may be more akin to ``normal" LSB dwarfs, but with unusually large sizes due perhaps to unusual initial conditions, such as high angular momentum content (e.g., \citealp{Amo16}) or a particularly bursty star formation history \citep{Chan18,DiC17}. The study of satellite galaxies in galaxy clusters also suggest that tidal stripping may contribute to the formation of at least some UDG-like dwarfs \citep{Car19, Sal20}. 

Our recent analysis of the GC systems for UDGs in the Coma cluster revealed large object-to-object variations in GC specific frequency, suggesting that objects belonging to this somewhat loosely defined class may not share a single formation channel \citep{Lim18}. Velocity dispersion measurements and stellar population studies similarly suggest that UDGs may have formed via multiple processes \cite[see, e.g.,][]{Zar17,Fer18,Tol18}. There is some evidence too that environment might play a role in their formation: i.e., a roughly linear relationship exists between host cluster mass and the total number of UDGs \citep{vanB17}, although the slope is still under debate \citep{Man18}. 

The many questions surrounding these puzzling objects --- which include even the appropriateness of the claim that they make up a unique and distinct galaxy class (see \citealp{Con18,Dan19}) --- stem, in large part, to the incomplete and heterogeneous datasets that have been used to find and study them. Ideally, deep, wide-field imaging that is sensitive to both LSB and ``normal" galaxies --- across a range of luminosity and environments --- is required to detect and study these objects, and to explore the mechanisms by which they formed and their relationship to ``normal" galaxies.

As the rich cluster of galaxies nearest to the Milky Way, the Virgo Cluster is an obvious target for a systematic study for all types of stellar systems including UDGs. The {\it Next Generation Virgo Cluster Survey} \cite[NGVS;][]{Fer12} is a powerful resource for discovering, studying and inter-comparing UDGs and normal galaxies in this benchmark cluster. Indeed, the NGVS has already been used to study a wide range of galaxy types in this environment. Previous papers in this series have examined the properties of structurally extreme galaxies in Virgo, including both compact \citep{Zha15,Gue15,Liu15,Liu16,Zha18} and extended systems (i.e., UDGs). For the latter population, previous NGVS papers have reported the initial discovery of such galaxies in Virgo \citep{Mih15}, their kinematics and dark matter content \citep{Tol18}, and their photometric and structural properties \citep{Cot20}. Other NGVS papers have examined the detailed properties of Virgo galaxies \citep{Fer20}, including distances \citep{Can18}, intrinsic shapes \citep{San16}, nuclei and nucleation rates \citep{Spe17,San19}, color-magnitude relations \citep{Roe17}, luminosity functions \citep{Fer16} and abundance matching analyses \citep{Gro15}.

This paper is structured as follows. In \S\ref{data}, we present an overview of the NGVS survey: its design, imaging materials and data products, including the catalog of Virgo Cluster member galaxies that forms the basis of our study. In \S\ref{results}, we describe our selection criteria for UDGs as well as their observed and derived properties, such as the luminosity function, structural parameters, spatial distribution, and globular cluster content. In \S\ref{discussion}, we discuss our findings in the context of UDG formation scenarios. We summarize our findings and outline directions for future work in \S\ref{summary}. In an Appendix, we present notes on the individual galaxies that satisfy some, or all, of our UDG selection criteria.

\section{Observations and Data Products}
\label{data}

\subsection{Galaxy Detection}

The NGVS is a deep imaging survey, in the $u^*g'r'i'z'$ bands, carried out with MegaCam on the Canada-France-Hawaii Telescope (CFHT) over six consecutive observing seasons from 2008 to 2013. The survey spans an area of 104~deg$^2$ (covered in 117 distinct NGVS pointings) contained within the virial radii of the Virgo A and Virgo B subclusters, which are centered on M87 and M49, respectively. Full details on the survey, including observing and data processing strategies, are available in \cite{Fer12}. A complete description of the data reduction and analysis procedures, including the galaxy catalog upon which this study is based, can be found in \cite{Fer20}. 

Briefly, an automated identification pipeline for candidate Virgo Cluster member galaxies --- optimized for the detection of low-mass, low surface brightness (LSB) systems that dominate the cluster population by numbers --- was developed using a training set of 404 visually selected and quantitatively vetted cluster members in the Virgo core region. Galaxies, identified over the entire cluster using this custom pipeline, were assigned a membership probability based on several pieces of information, including location within the surface brightness vs. isophotal area plane; photometric redshifts from LePhare \citep{Arn99,Ilb06}; goodness-of-fit statistics and model residuals; and the probability of a spurious detection arising from blends or image artifacts. Following a final series of visual checks, candidates were assigned membership classifications ranging from 0 to 5. In this analysis, we restrict ourselves to the 3689 galaxy candidates classified as types 0, 1 or 2: i.e., ``certain", ``likely" or ``possible" Virgo cluster members. Note that our sample of UDGs includes no Class 2 objects, and roughly equal numbers of Class 0 and  Class 1 types. For reference, the mean membership probabilities for these classes are $84\pm23\%$ (class 0) and $77\pm21\%$ (class 1; see Figure~12 of \citealt{Fer20}).

\subsection{Estimation of Photometric and Structural Parameters}

Photometric and structural parameters were measured for each galaxy using up to three different techniques, depending on the galaxy magnitude and surface brightness. First, for most galaxies brighter than $g' \simeq 17$, an isophotal analysis was carried out using a custom software package, {\tt Typhon} \citep{Fer20}, that is built around the {\tt IRAF ELLIPSE} and related tasks, followed by parametric fitting of the one-dimensional surface brightness profile using one- or two-component S\'ersic models (with the possible inclusion of a central point source for nucleated galaxies). Second, basic global parameters (e.g., integrated magnitudes, mean and effective surface brightnesses, half-light radii) were measured non-parametrically using a curve-of-growth analysis, including an iterative background estimation. Finally, galaxies fainter than $g' \sim 16$ were also fitted in the image plane with GALFIT \citep{Pen02}, with assumed S\'ersic profiles (with, or without, point source nuclei). In our study, which relies entirely on global parameters for UDGs and normal galaxies, we use parameters from the {\tt Typhon} analysis whenever available; otherwise, we rely on GALFIT parameters derived from two-dimensional fitting of the images. In \citet{Fer20}, it is shown that these techniques yield parameters that are in good statistical agreement for the majority of galaxies. Nevertheless, one should bear in mind that the subjects of this paper, UDGs, are among the faintest and most diffuse galaxies in the NGVS catalog, and thus present special challenges for the measurement of photometric and structural parameters. 

To gauge survey completeness, an extensive series of simulations were carried out in which artificial galaxies --- convolved with the appropriate point spread function and with appropriate amounts of noise added --- were injected into the core region frames, and then recovered using the same procedure employed to build the galaxy catalog. In all, 182,500 simulated galaxies were generated, equally divided among the $u{^*}g'r'i'z'$ filters, under the assumption that their intrinsic light distributions follow S\'ersic profiles. Simulated galaxies were randomly generated so as to populate (and extend beyond) the scaling relations expected for Virgo members, with S\'ersic indices, effective radii and total magnitudes in the range $0.4 \le n \le 2.5$, $0\farcs9 < R_e < 53\arcsec~$ ($75 \le R_e \le 4200$~pc), and $16.0 \le g' \le 25.2$ mag. Completeness contours for each scaling relation were then derived, with a 50\% completeness limit in magnitude of $g' \simeq 21.5$~mag ($M_g \simeq -9.6$~mag). For a thorough discussion of catalog completeness and the reliability of our photometric and structural parameters, the reader is referred to \cite{Fer20}, who describe the completeness tests in detail, and \cite{Cot20}, who compare the measured NGVS photometric and structural parameters to those in the literature for previously cataloged Virgo galaxies.

\begin{figure*} 
\epsscale{1.0}
\plotone{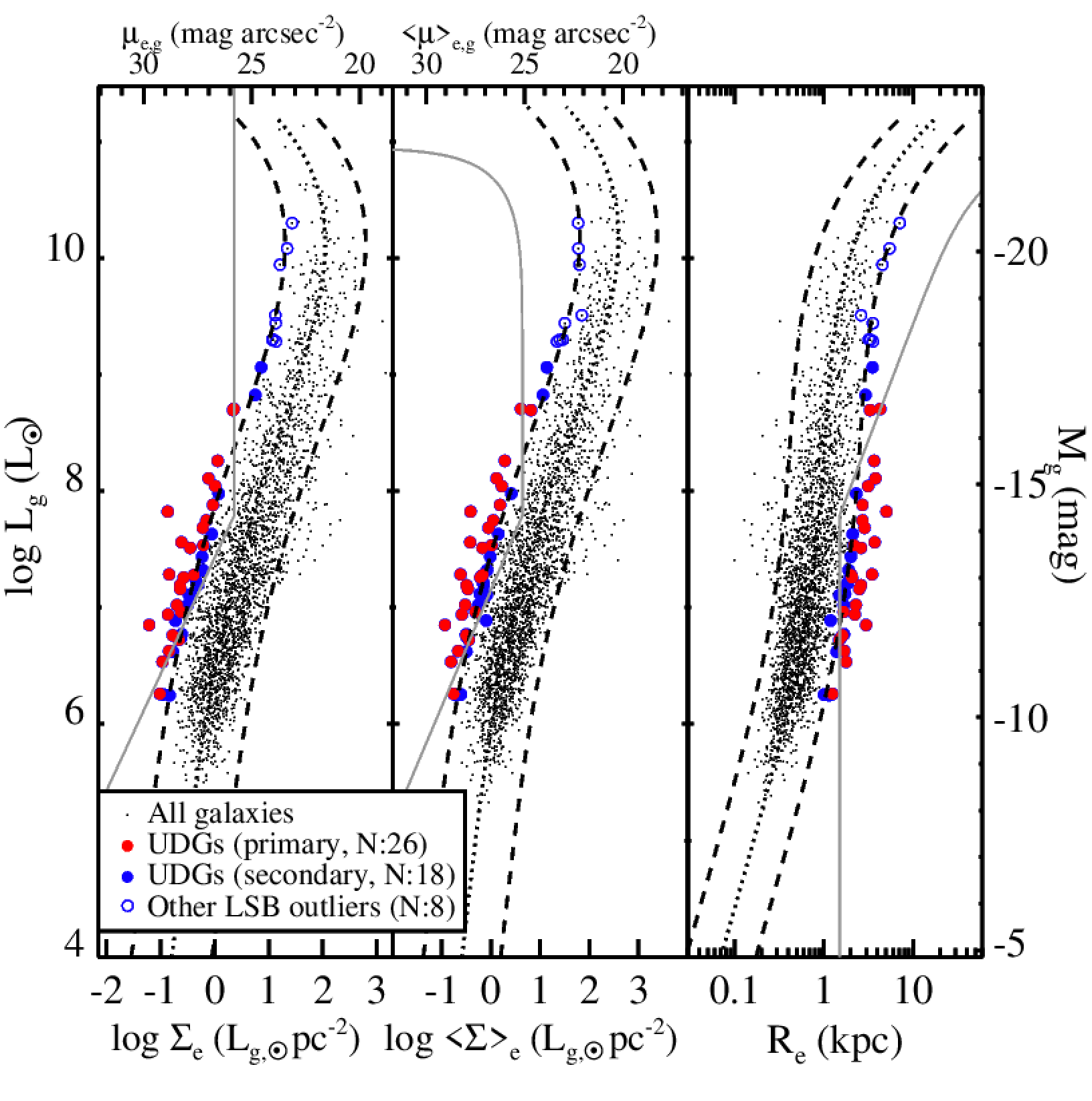}
\caption{Scaling relations (in the $g'$ band) between luminosity, effective radius and surface brightness, and mean effective surface brightness for 3689 galaxies in the Virgo cluster (small black points). The filled red and blue circles show our UDGs (i.e., the {\tt primary} and {\tt secondary} samples, respectively; see \S\ref{selection}). Eight bright spiral galaxies that were initially selected as LSB outliers in our secondary sample, but discarded from our analysis, are also shown (open blue circles). The dotted and dashed curves show the mean scaling scaling relations and their $2.5\sigma$ confidence limits, respectively. The gray solid curve in each panel shows the UDG definitions adopted by \citet{vanD15}. 
\label{udgsel}}
\end{figure*}

\section{Results}
\label{results}

\subsection{Identification of UDG Candidates}
\label{selection}

Large and LSB dwarf galaxies have been known for some time (see, e.g., Table XIV and Appendix C-4 of \citealt{Bin85}), but the notion that UDGs comprise a new and distinct galaxy class was introduced by \citet{vanD15}. In their Dragonfly survey of the Coma Cluster, these authors defined UDGs as those galaxies with central $g'$-band surface brightnesses below $\mu_0 = 24$ mag~arcsec$^{-2}$ and effective radii larger than $R_e = 1.5$ kpc. It was subsequently noted that this size criterion is close to the limit set by the resolution of Dragonfly at the distance of Coma: i.e., Full Width at Half Maximum (FWHM) = 6\arcsec~ $\sim$ 3~kpc. Other studies (e.g., \citealt{Mih15,Kod15,Yag16,vanB16,vanB17}) have used different criteria (e.g., different size cuts and/or different $\mu_0$ or $\langle\mu(R,R_e)\rangle$ limits), which has led to some confusion on the properties of UDGs as a class. It is clear that {\it we require new classification criteria that are based on the physical properties of galaxies and largely independent of the classification material.}

\begin{figure} 
\epsscale{1.2}
\plotone{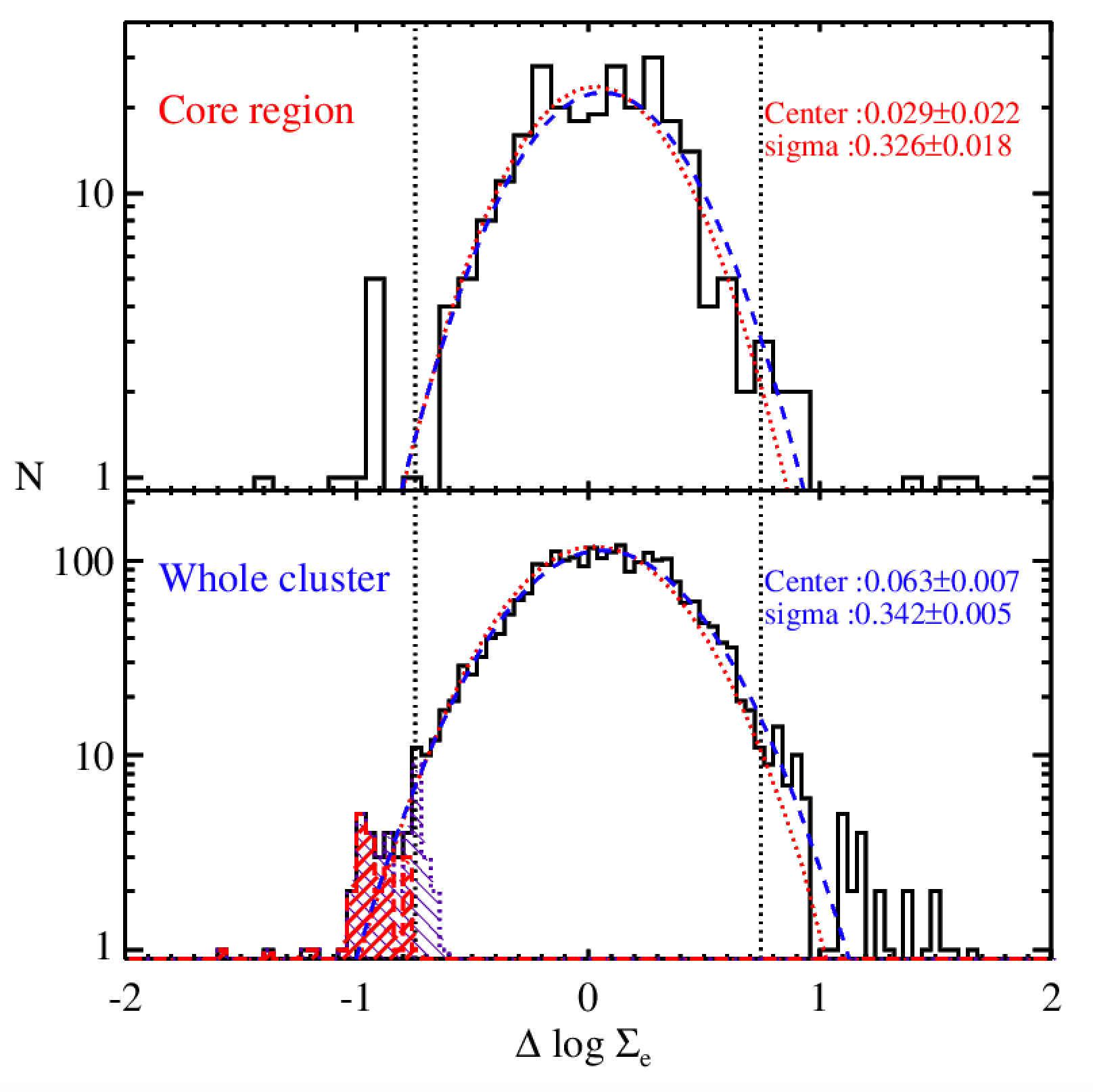}
\caption{Deviations from the mean relation between luminosity and effective surface brightness, $\Sigma_e$. The vertical lines show 2.5$\sigma$ deviations from the mean which we used in UDG definition (the positions of dashed lines in Figure 1), while the dotted and dashed curves show Gaussian fits to the distributions of galaxies in the core region and whole cluster, respectively. Note that the dashed curve in the upper panel and the dotted curve in the lower panel have been renormalized. The solid histogram in the upper panel shows galaxies in the core region, while the solid histogram in the lower panel shows all cluster member galaxies from the NGVS. The dashed red histogram in the lower panel shows the UDG {\tt primary} sample; results for the combined {\tt primary and secondary} samples are shown in purple. 
\label{dSBe}}
\end{figure}

Our selection criteria for UDGs is based on scaling relations for 404 galaxies in Virgo's core region, which is a $\simeq$ 4 deg$^2$ region roughly centered on M87, the central galaxy in Virgo's A sub-cluster \citep{Cot20}. This study used NGVS data to examine the relationships between a variety of photometric and structural parameters, including size, surface brightness, luminosity, color, axial ratio and concentration. In Figure \ref{udgsel}, we show scaling relations, plotted as a function of $g$-band magnitude, for the full NGVS catalog of 3689 certain, likely or possible Virgo Cluster members \citep{Fer20}. From left to right, the three panels of this figure show: (1) surface brightness measured at the effective radius, $\Sigma_{e}$; (2) mean surface brightness measured within the effective radius, $ \langle \Sigma \rangle_{e}$; and (3) effective radius, $R_e$, which is defined as the radius containing half of the total luminosity.  The dotted curve in each panel of Figure~\ref{udgsel} shows a fourth order polynomial, as recorded in Table~8 of \cite{Cot20}, which was obtained by maximum likelihood fitting to the observed scaling relations for galaxies in the Virgo core region. These polynomials were fitted over the luminosity range $10^5 \le {\cal L}_g/{\cal L}_{g,\odot} \le 10^{11}$ ($-8 \lesssim M_g \lesssim -22.4$~mag) and account for incompleteness by modifying the likelihood function with weights that are inversely proportional to the completeness functions derived from the galaxy simulations described in \S\ref{data} (\citealt{Fer20}).

\begin{figure} 
\epsscale{1.2}
\plotone{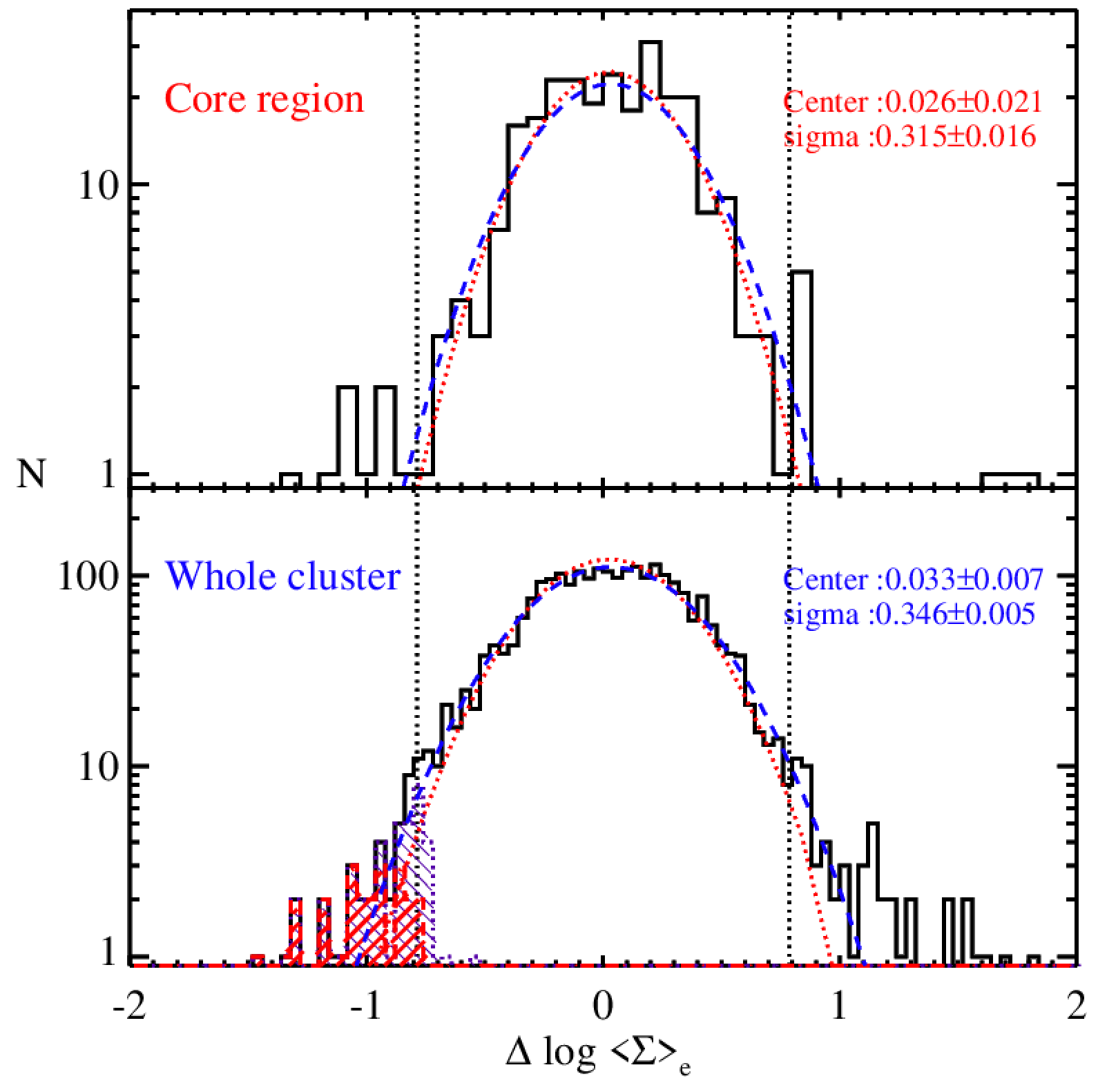}
\caption{Same as Figure \ref{dSBe}, but showing the deviations from the mean relation between luminosity and mean effective surface brightness, $\langle\Sigma\rangle_e$.  
\label{dmSBe}}
\end{figure}

\begin{figure} 
\epsscale{1.2}
\plotone{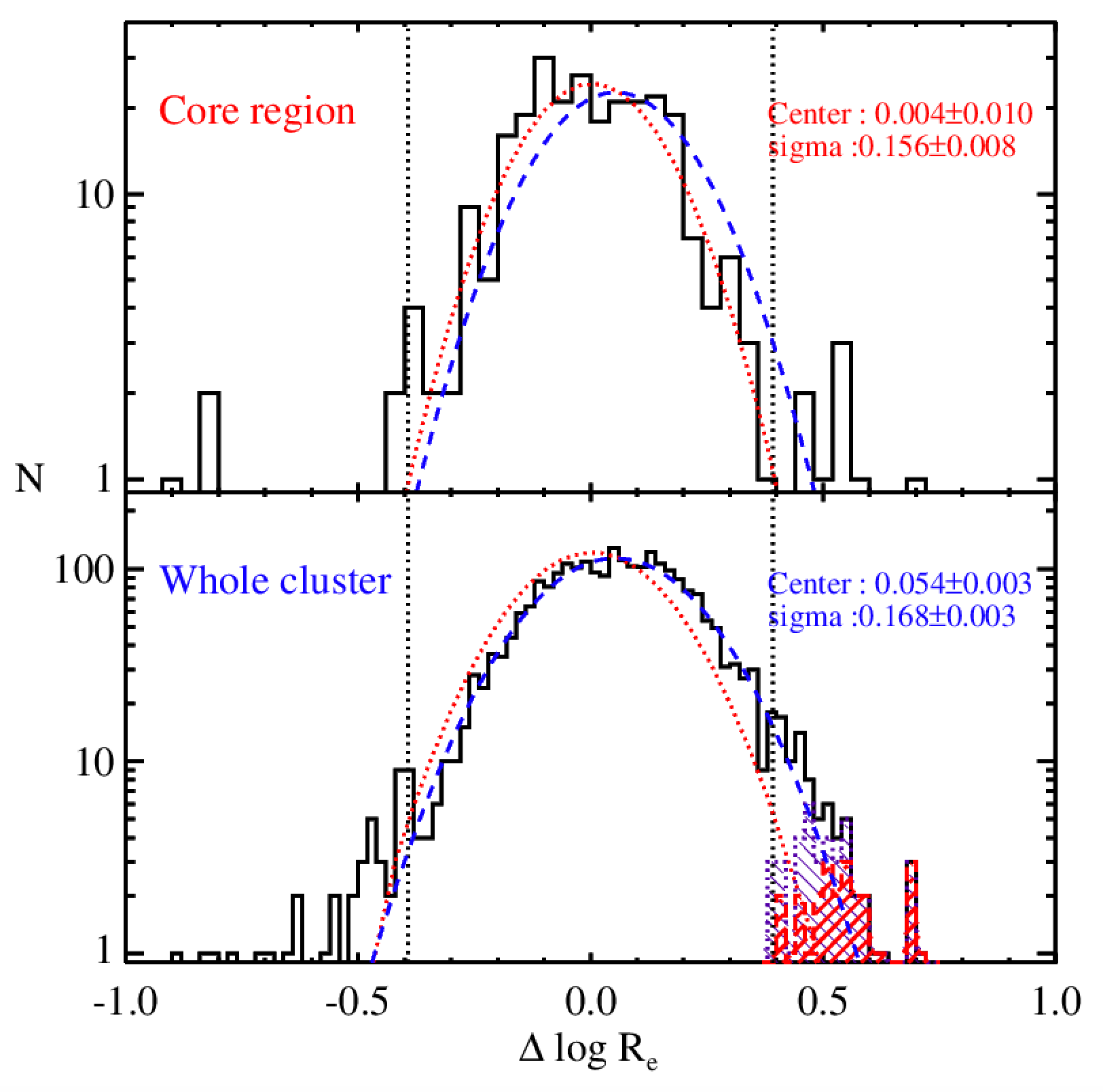}
\caption{Same as Figure \ref{dSBe}, but showing the deviations from the mean relation between luminosity and effective radius, $R_e$.  
\label{dRe}}
\end{figure}

We have opted to base our selection criteria on galaxies in the cluster core because this region has the full set of information needed to establish cluster membership and determine scaling relations. The relations themselves are quite similar to those found for galaxies over the whole cluster and, indeed, Figures~\ref{udgsel}--\ref{dRe} demonstrate that these relations provide an adequate representation of the scaling relations for the full sample of galaxies. To identify UDG candidates in a systematic way, we wish to select galaxies that have unusually large size, and/or unusually low surface brightness, at fixed luminosity. We thus use the scatter about the fitted polynomials in the Virgo core region to identify outliers in these diagrams. The dashed curves in each panel show the $\pm2.5\sigma$ band that brackets each polynomial. The standard deviation, $\sigma$, in each case has been determined using the sample of 404 central galaxies (i.e., given the limited sample size, we make no attempt to quantify any variation in $\sigma$ with luminosity). 

We are now in a position to select UDG candidates using these relations. For galaxies in the core region, the distributions of deviations from the mean relations are shown in the upper panels of Figures \ref{dSBe}, \ref{dmSBe} and \ref{dRe}. In these distributions, particularly those involving effective surface brightness, there may be a gap close to $2.5 \sigma$ from the mean relation. We henceforth use this condition for identifying LSB outliers, defining our {\tt primary sample} of UDGs to be those systems that deviate --- towards large size or low surface brightness --- by more than $2.5 \sigma$ in {\it each of the three scaling relations}. 

While these requirements should produce a robust sample of UDGs, it is likely that some bonafide LSB systems may be missed by these stringent selection criteria. We therefore consider an augmented sample of galaxies that satisfy only two, or one, of these criteria. Therefore, in the analysis that follows, we rely on two UDG samples:

\begin{itemize}
	\item[1.] {\tt Primary (26 UDGs):} This sample is made up of the 26 galaxies that deviate by at least $2.5 \sigma$ in each of the three defining scaling relations: i.e., $L$-$R_e$, $L$-$\mu_e$ and $L$-$\langle\mu\rangle_e$. This sample has the benefit of {\it high purity} but may exclude some LSB objects that do not satisfy all selection criteria.
	\item[2.] {\tt Secondary (18 UDGs):} This sample is defined by starting with the 26 galaxies that deviate by at least $2.5 \sigma$ in only one, or two, of the scaling relations. Beginning with this sample, we have excluded eight bright (and face-on) spiral galaxies with $g$-band luminosities greater than $\sim$10$^{9.25}$ L$_{g,\odot}$, leaving us with 18 additional UDGs. The combined primary and secondary samples (28 + 18 = 44 objects) has the advantage of {\it high completeness}. 
\end{itemize}

In an appendix, we present more information on these samples, including tabulated parameters, color images, and detailed notes on individual objects. Meanwhile, Figure~\ref{udgsel} shows the distribution of our UDGs in each of the three scaling relations. Objects that belong to our {\tt primary sample} are shown as filled red circles in each panel while the 18 UDGs from our {\tt secondary sample} are shown as filled blue circles. The eight, bright spiral galaxies that were initially selected and discarded from our secondary sample are shown as open blue circles.

Table \ref{tbl:udgcat} lists information on the 26 UDGs that belong to our {\tt primary} sample (see the appendix for information on the {\tt secondary} sample). From left to right, the columns record the object name, right ascension and declination, magnitude, effective radius, effective surface brightness and mean effective surface brightness (all measured in the $g'$ band). The final two columns report previous names, if applicable, and the official object identification from the NGVS. 

\begin{figure} 
\epsscale{1.18}
\plotone{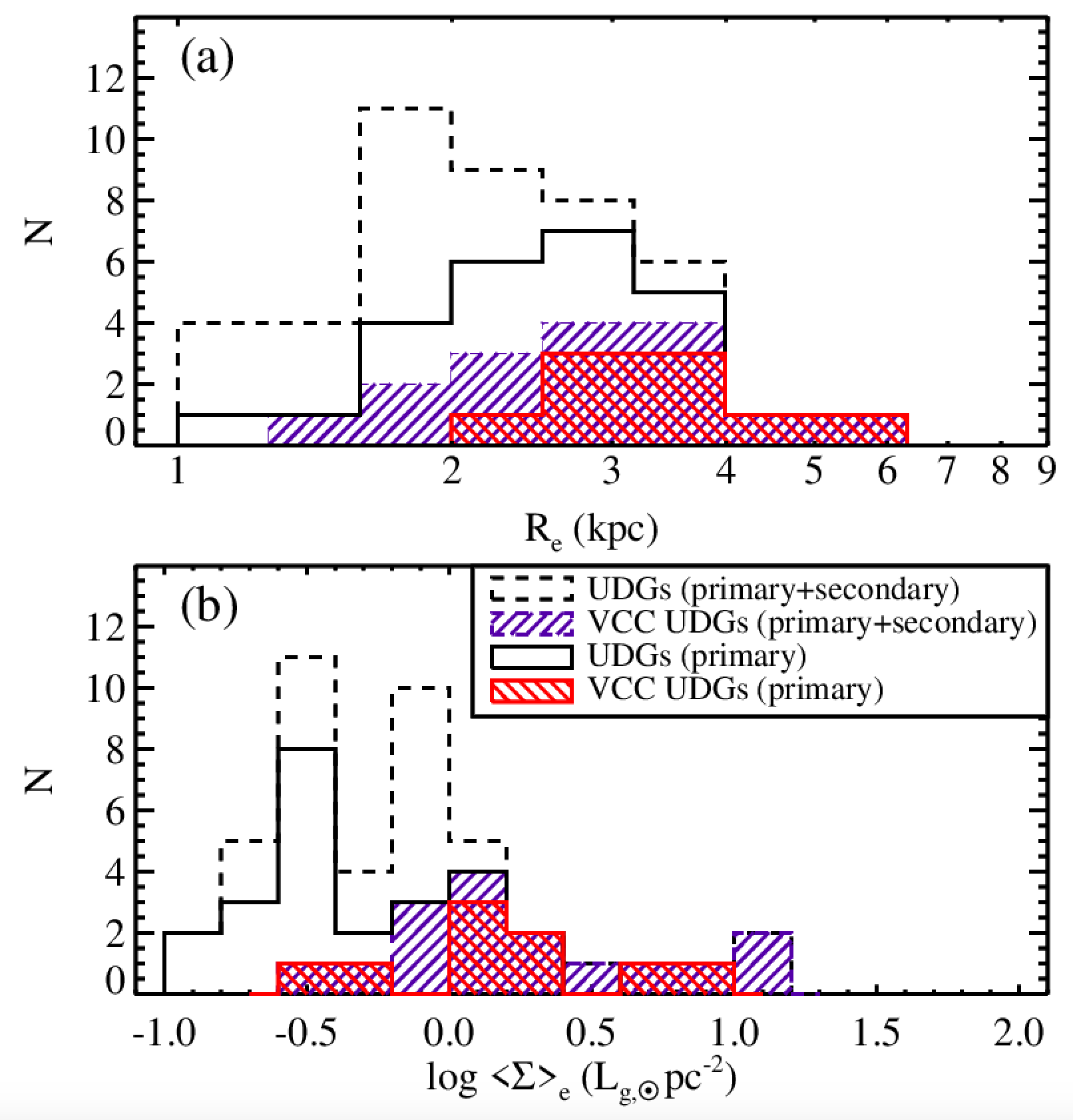}
\caption{Effective radii and surface brightness distributions of UDGs. {\it Panel (a)}. Distribution of UDG effective radii. The solid and dashed black histograms show distributions for the {\tt primary} and combined {\tt primary} and {\tt secondary} UDG samples, respectively. Hatched versions of these histograms show the subset of UDGs that appear in the VCC catalog of \citet{Bin85}. {\it Panel (b)}. The distribution of mean surface brightness within effective radius for UDGs. All histograms are the same as in the upper panel.  
\label{ngvsvcc}}
\end{figure}

Exactly half (13 $\simeq 50\%$) of the 26 galaxies in Table~\ref{tbl:udgcat} were previously cataloged, mostly in the Virgo Cluster Catalog (VCC) of \citet[but see also \citealt{Rea83}]{Bin85}. This is also true of the combined {\tt primary} and {\tt secondary} samples, where 21/44 $\simeq$ 48\% of the  UDGs are previously catalogued galaxies (though not necessarily identified in the past as extreme LSB systems). In Figure \ref{ngvsvcc}, we show distributions for the effective radius and mean surface brightness for our sample (upper and lower panels, respectively). For reference, we highlight the subsets of UDGs that were previously cataloged by \citet{Bin85}, the most comprehensive catalog of Virgo galaxies prior to the NGVS. As noted above, roughly half of these UDGs were previously cataloged galaxies, although the NGVS clearly excels in the detection of the smallest, faintest and lowest surface brightness UDGs.

\begin{figure*} 
\epsscale{1.2}
\plotone{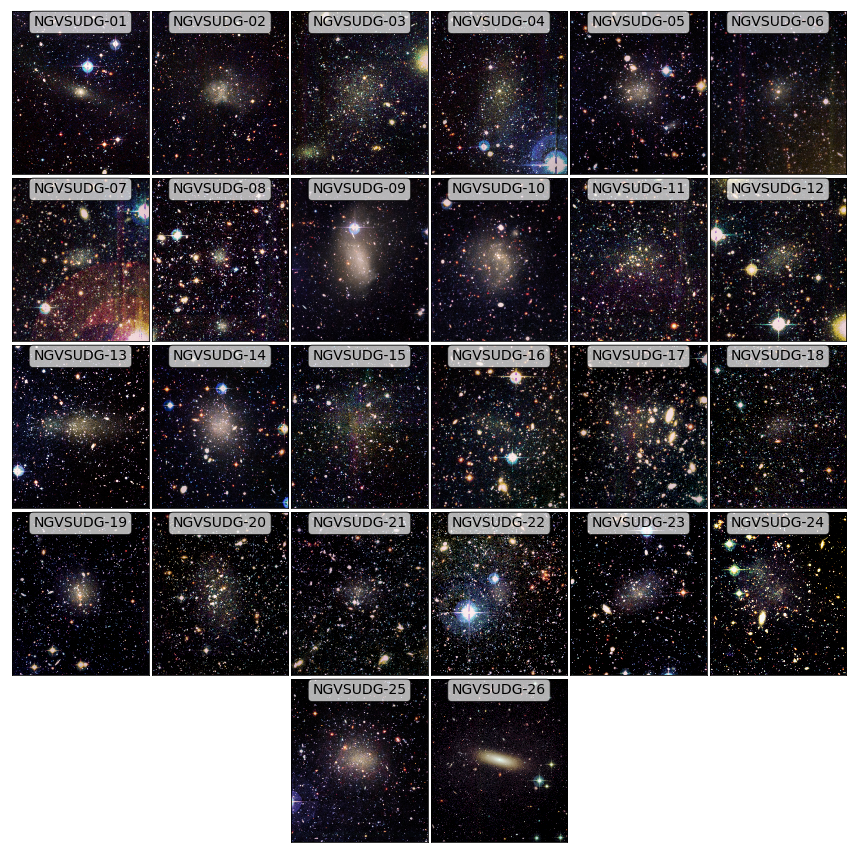}
\caption{Thumbnail images of UDGs from our {\tt primary} sample. The field of view is $4\farcm4 \times 5\farcm3$. North is up and East is left. The color is constructed from $u^*gi$ images. To enhance LSB features, all images have been smoothed using a Gaussian kernel with $\sigma = 2.5$~pixels (${\rm FWHM} = 1\farcs1)$. The ID of each UDG is noted at the top of its image. \label{udgthumb_primary}}
\end{figure*}

Figure \ref{udgthumb_primary} shows thumbnail images for UDGs belonging to our {\tt primary sample}. Several objects have elongated shapes that may point to tidal effects but others have a more circular appearance (see Appendix B for detailed notes on individual galaxies and \S\ref{morphology} for a more complete discussion of UDG morphologies and their implications). 

We can now compare our selection methodology to that of \citet{vanD15}. The solid gray curves in Figure~\ref{udgsel} show their criteria. Note that these relations exhibit a break just below $L_g \sim 10^8$ $L_{g,\odot}$. This is a consequence of the fact that the original criteria were specified in terms of effective radius and {\it central} surface brightness, for an assumed exponential profile.\footnote{For the NGVS, galaxies are fitted using a more generalized S\'ersic model, which is flexible enough to capture the change in galaxy concentration with luminosity and mass.} For the most part, the selection criteria are in reasonable agreement although it is clear that the \citet{vanD15} definition does include some ``normal" galaxies around luminosities of $L_g \sim 10^8 L_{g,\odot}$. Our objective method initially selects some luminous late-type galaxies in the {\tt secondary} sample, but these objects are not present in the {\tt primary} sample. In our UDG selection criteria, there is a room for faint UDGs to have effective radii smaller than $R_e = 1.5$ kpc, but we find only a single UDG in the {\tt primary} sample (NGVSUDG-08 = NGVSJ12:27:15.75+13:26:56.1) to have an effective radius less than $R_e=1.5$ kpc. The small number of outliers among faint galaxies is possibly due to the onset of incompleteness in the survey.

Returning to Figures~\ref{dSBe}, \ref{dmSBe}, and \ref{dRe}, the bottom panels of these figures show the deviations from the mean scaling relations presented in Figure~\ref{udgsel}. For the residual surface brightness distributions, the fitted Gaussians have means and standard deviations that are consistent, within their respective errors, between the core region and entire cluster. The residual effective radius distributions are slightly broader for the full sample of galaxies than those in the core region, evidence that low-mass galaxies in low-density environments have larger sizes. 

It should be noted that we also find a number of compact galaxies located on the {\it opposite} side of UDGs in the scaling relation sequences. These relatively rare objects (i.e., cEs, BCDs) are distinct from ultra-compact dwarfs \citep[see][]{Liu15}, but are found throughout the cluster and span a wide range in luminosity and color. We will examine these objects in detail in a future paper.

\subsection{Luminosity Function}

\begin{figure} 
\epsscale{1.15}
\plotone{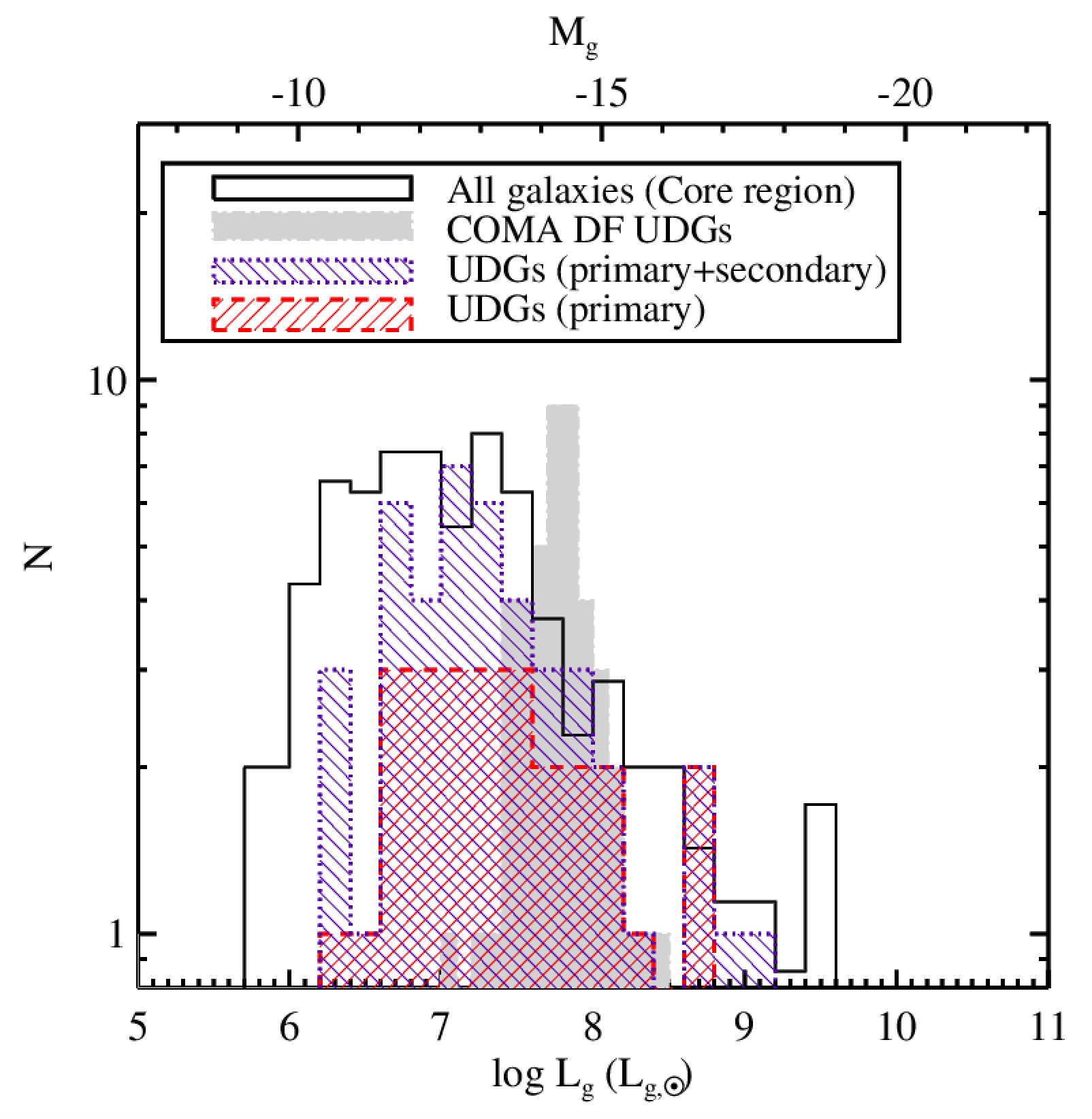}
\caption{The $g$-band luminosity function for Virgo UDGs. The red dashed histogram shows the luminosity function for our {\tt primary} sample of 26 UDGs; results for the combined {\tt primary} and {\tt secondary} samples (44 objects) are shown as the purple dotted histogram. The luminosity function for galaxies in the Virgo core region is shown as the solid histogram (after renormalizing). For comparison, the luminosity function of Coma UDGs from \citet{vanD15} is shown by the gray histogram. 
\label{udglf}}
\end{figure}

Figure \ref{udglf} shows luminosity functions for our UDG samples. As explained in \S\ref{selection}, we exclude from this analysis the eight, bright, late-type systems whose luminosities range from $10^{9.3}$ to $10^{10.3}$ $L_{g,\odot}$. The brightest of the 26 UDGs in the {\tt primary} sample has a luminosity of $\sim\!\!\!10^{8.7} L_{g\odot}$ --- comparable to that of the brightest of the 18 UDGs in the {\tt secondary} sample, which has $\sim\!\!10^{9.1} L_{g,\odot}$. The faintest objects in either sample have $\sim\!\!10^{6.2} L_{g,\odot}$, which is slightly brighter than the detection limit of the survey. The luminosities of our faintest UDGs are well below those of the UDGs discovered in the Coma cluster by Dragonfly \citep{vanD15}, a reflection of the depth and spatial resolution afforded by NGVS imaging. 

Broadly speaking, the luminosity distribution of the combined {\tt primary} and {\tt secondary} sample is fairly similar to that of ``normal" Virgo Cluster galaxies \cite[see][]{Fer16}. The luminosity function of the {\tt primary} sample alone appears flatter than that of  ``normal" galaxies although the relatively small number of galaxies (26) limits our ability to draw firm conclusions. We caution that, for either UDG sample, selection effects can be significant given the faint and diffuse nature of these galaxies (see \citealt{Fer20})

\begin{figure} 
\epsscale{1.15}
\plotone{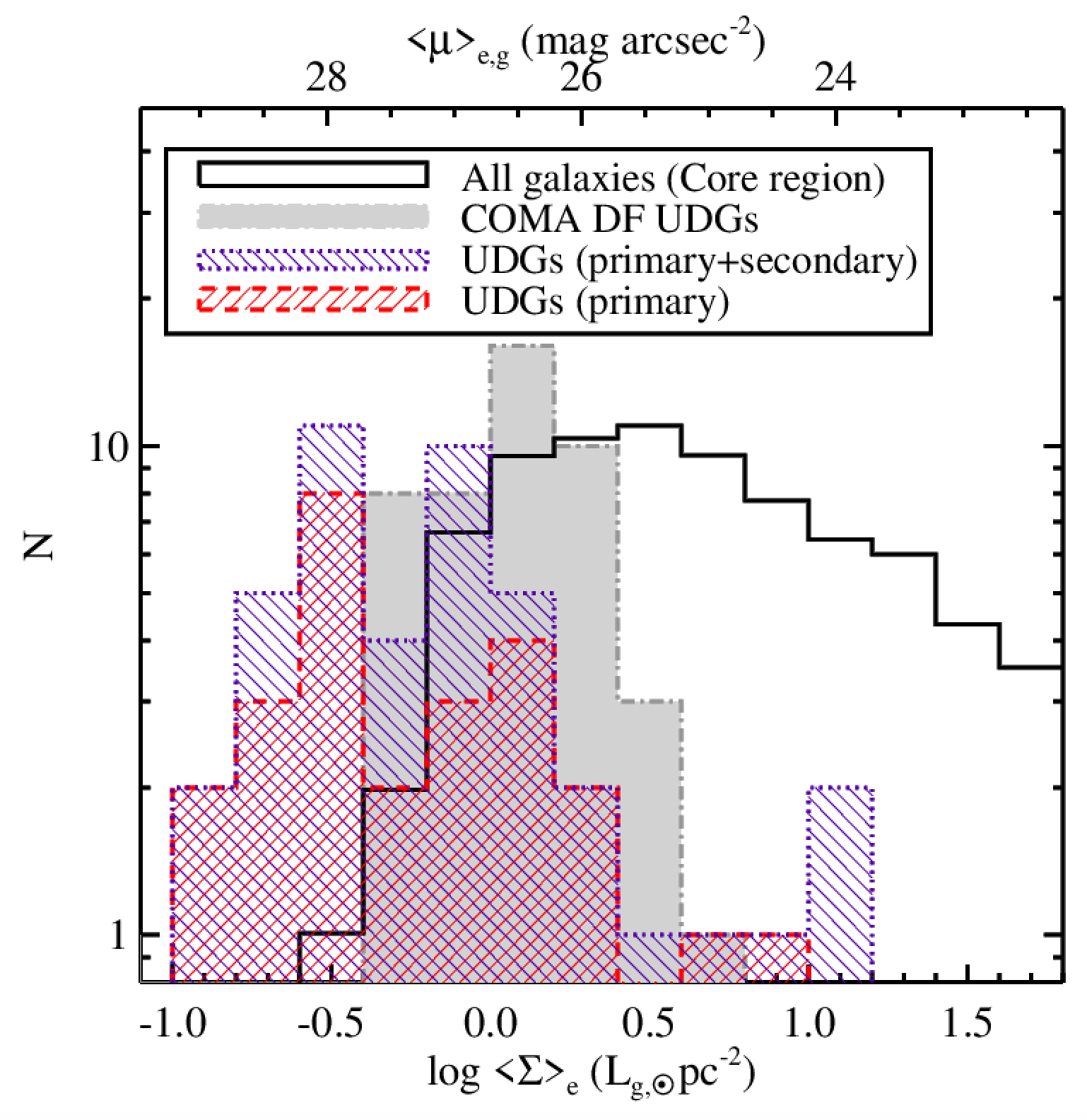}
\caption{The distribution of mean effective surface brightness for UDGs. Symbols and notations are the same as those in Figure \ref{udglf}.  
\label{udgmSBe}}
\end{figure}

Figure~\ref{udgmSBe} shows the distribution of effective surface brightness, $\langle \Sigma \rangle _{e}$, for our  UDGs.  The UDGs belonging to our {\tt primary} sample span a range of $10^{-1.0} \lesssim \langle\Sigma\rangle_e \lesssim 10^{1.0} L_{\odot}~{\rm pc^{-2}}$. Overall, the number of UDGs increases with decreasing surface brightness until $\langle\Sigma\rangle_{g,R_e} \sim 10^{-0.6} L_{g,\odot}{\rm pc^{-2}}$ (which is equivalent to $\langle\mu\rangle_e \sim 28~ {\rm mag~arcsec^{-2}}$ in the $g$ band).  Naturally, about half of the Virgo UDGs are fainter than the mean surface brightness of $\langle\Sigma\rangle_e \sim 10^{-0.3} L_{g,\odot}{\rm pc^{-2}}$ (or $\langle\mu\rangle_e \sim 27.5 \, {\rm mag \, arcsec^{-2}}$ in the $g$ band). This corresponds to the surface brightness of the faintest Coma Dragonfly UDGs \citep{vanD15}, and is significantly fainter than most other UDG surveys (e.g., \citealp{vanB16,Rom17,Man18}). If Virgo is representative of these other environments, then this would suggest that other clusters may contain significantly more very low surface brightness UDGs than currently cataloged.
Although we found a significant population of very low surface brightness UDGs, the number of UDGs in the Virgo cluster is consistent with the expected number from the \citet{vanB16} relation between halo mass and number of UDGs when we use a survey limit similar to that in previous studies.

\subsection{Spatial Distribution}

\begin{figure} 
\epsscale{1.15}
\plotone{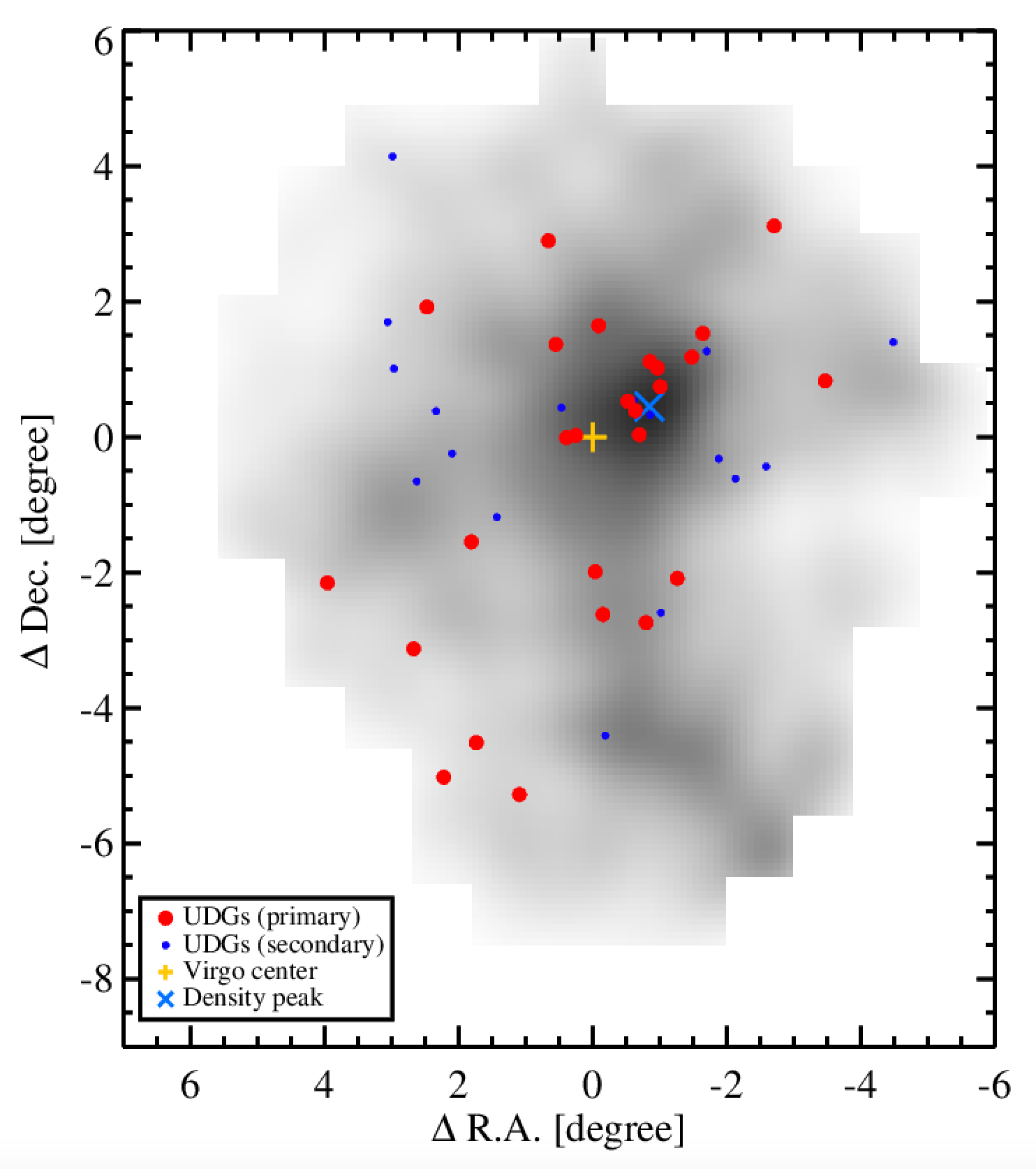}
\caption{The spatial distribution of Virgo UDGs within the NGVS footprint. The grayscale map shows the surface number density of $\sim$3689 Virgo cluster member galaxies from the NGVS. Red and blue symbols show UDGs, {\tt primary }and {\tt secondary} samples, respectively. The orange and light blue crosses show the center of the Virgo cluster and the peak of the galaxy density distribution, respectively. 
\label{udgspa}}
\end{figure}

\begin{figure} 
\epsscale{1.15}
\plotone{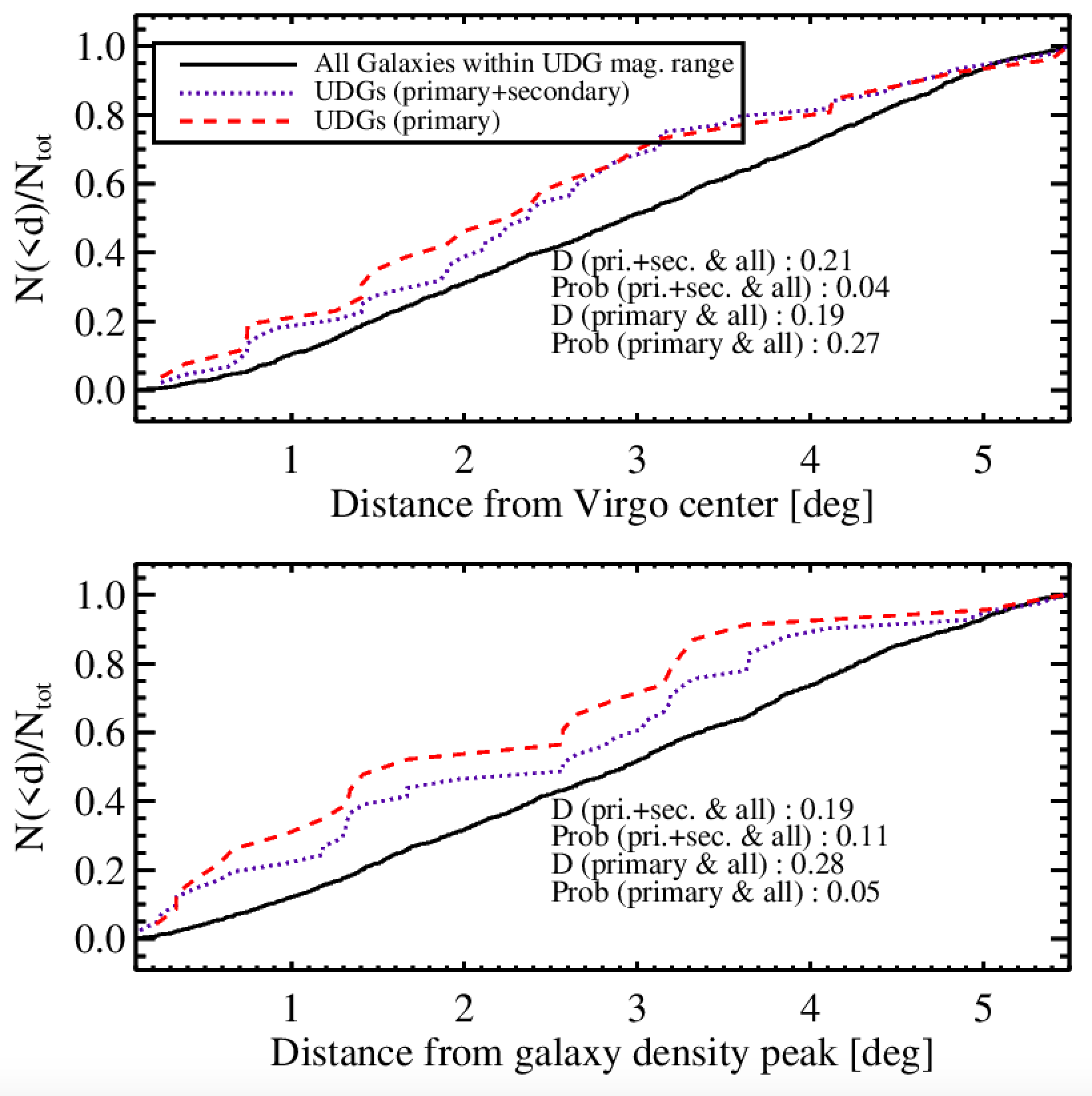}
\caption{The cumulative distribution of cluster-centric radii for Virgo Cluster UDGs. The upper panel shows the distribution of distances from the Virgo cluster center (i.e., the centre of M87). The lower panel shows the distribution of distances from the location of the galaxy peak density (see Figure~\ref{udgspa}). In both panels, the red dashed and purple dotted curves show the distributions for the {\tt primary}, and combined {\tt primary} and {\tt secondary} samples, respectively. Results from KS tests are summarized in each panel. {\it D} is the maximum deviation between the two distributions, and {\it prob} is the significance level of the KS statistic. A small {\it prob} value means more significantly different distributions.
\label{cumudg}}
\end{figure}

The distribution of UDGs within the cluster may hold clues to their origin, and we begin by noting that the cluster core appears to be overabundant in UDGs. In fact, 10 of the 44 galaxies (23\%) in our combined {\tt primary} and {\tt secondary} samples are found in the central 4~deg$^2$. While we cannot rule out the possibility that some of these candidates are being seen in projection against the cluster center, the enhancement is likely real as this region represents $\lesssim 4\%$ of the survey area. We shall return to this issue in \S\ref{spatial}.

Figure~\ref{udgspa} shows the distribution of UDGs within the Virgo Cluster. Objects belonging to the {\tt primary} and {\tt secondary} samples are are shown separately. The underlying gray scale map shows the surface density of the 3689 certain or probable member galaxies from the NGVS. The orange cross shows the location of M87 --- the center of the Virgo A sub-cluster and the galaxy that has traditionally been taken to mark the center of Virgo. This figure shows that the UDGs are distributed over the entirety of the cluster, yet concentrated toward the cluster center. Additionally, it appears the UDGs are offset from both M87 and from the centroid of the galaxy population (marked as a light blue cross in this figure), although they are more closely associated with the latter. We note that the offset of the galaxy density centroid from M87 is in the direction of the infalling M86 group, and the spatial distribution of UDGs is also offset in this direction.

To examine the concentration of the UDG population in more detail, we compare the cumulative distributions of UDGs and other galaxies in Figure~\ref{cumudg}. The upper and lower panels of this figure show the cumulative distribution of distances from M87 and the galaxy centroid, respectively. Whichever center is used, the UDGs --- from both the {\tt primary} and {\tt secondary} samples --- are found to be more centrally concentrated than other cluster galaxies. This is also true if we use the definition of UDGs from \citet{vanD15}. The results of our Kolmogorov-Smirnov (KS) tests show that radial distributions (centered on the galaxy density peak) for all galaxies and the {\tt primary} sample differ with $95\%$ probability. This finding is notable, as it differs from some previous studies that found UDGs in other clusters to be less centrally concentrated than normal galaxies (e.g., \citealp{vanB16,Tru17,Man18}).

\begin{figure} 
\epsscale{1.15}
\plotone{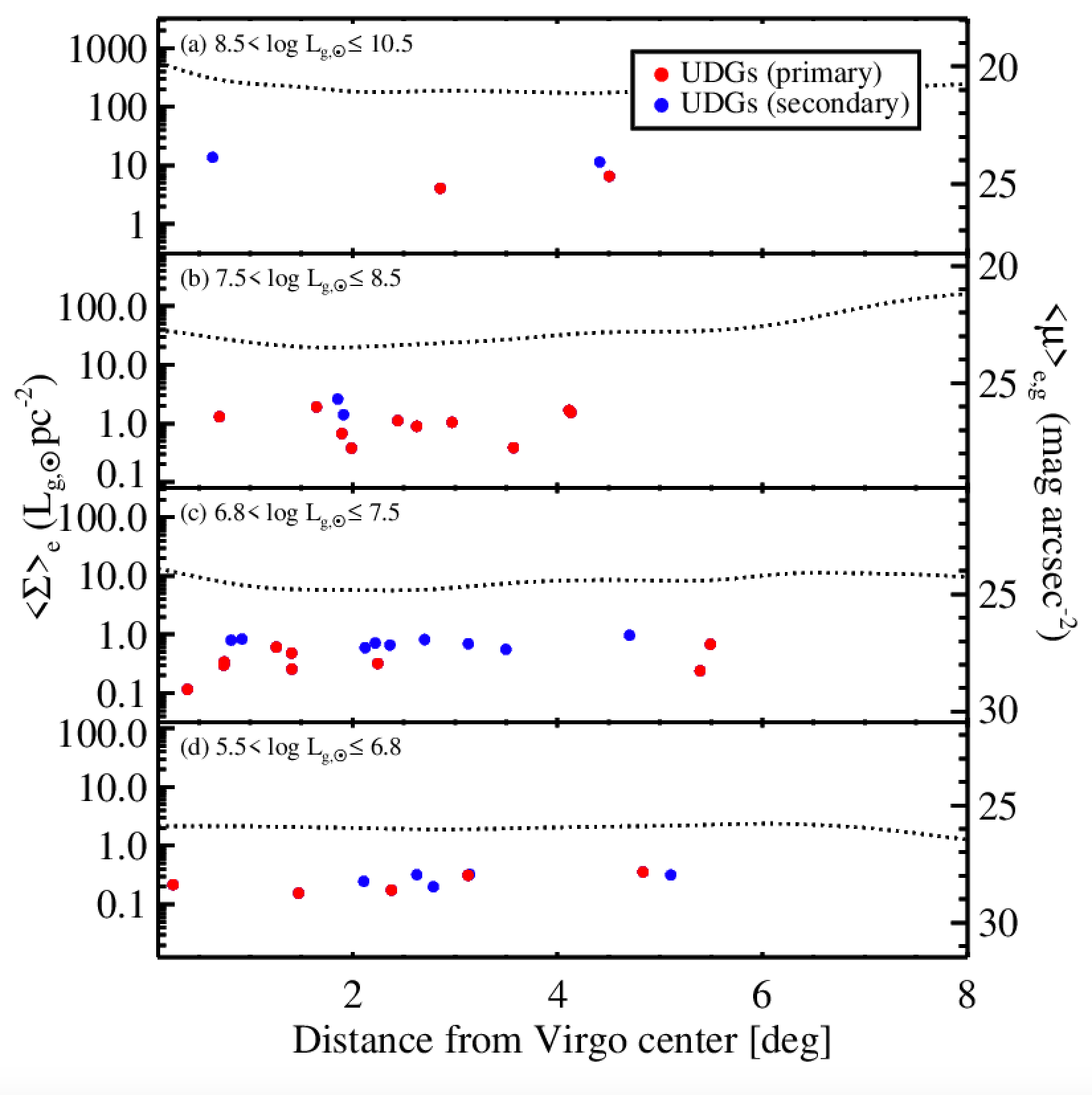}
\caption{Mean effective surface brightness of UDGs, in four different luminosity bins, plotted as a function of distance from the Virgo Cluster center. The black dotted curves show the mean trends for normal galaxies. The red and blue circles show UDGs from our {\tt primary} and {\tt secondary} samples, respectively.  
\label{dismSBe}}
\end{figure}

Figure~\ref{dismSBe} shows how the effective surface brightness of UDGs varies with distance from the cluster center. The four panels show the trends for galaxies divided into four bins of luminosity that decease from the top to bottom panels. The symbols are the same as in previous figures, and each panel includes a dotted curve that indicates the mean behavior of other cluster members. As expected, the UDGs fall well below the mean surface brightness of other galaxies at all radii. Interestingly, there is no apparent dependence of $\langle\Sigma\rangle_{e}$ on distance in each luminosity bin. However, we find few UDGs with $\langle\Sigma\rangle_e \gtrsim 1$ $L_{g,\odot}{\rm pc}^{-2}$ in the inner regions of the cluster although such objects should be detectable, if present. Indeed, the UDGs we do find in the central region are often fainter than the surface brightness limits of previous surveys (e.g., \citealp{vanB16,Tru17,Man18}). Although the Virgo B sub-cluster complicates the use of clustercentric distance as a proxy for environment, the effect is minimal because most UDGs in the Virgo B sub-cluster are found outside the dense sub-cluster core (see Figure \ref{udgspa}).

\subsection{Globular Cluster Systems}
At first, the diffuse nature of UDGs seems at odds with the formation of  massive star clusters, as the latter require environments with a high density of star formation to form \citep[cf.,][]{Kru14,Kru15}. However, many UDGs harbor significant populations of GCs \citep[e.g.,][]{Pen16,vanD17,Amo18,Lim18}.
We have used the NGVS data to examine the GC content of UDGs belonging to our {\tt primary} and {\tt secondary} sample, selecting high probability GC candidates on the basis of their $u^*g'i'$ colors and concentration indices \citep[see][]{Mun14,Dur14}. 

GCs at the distance of Virgo are nearly point sources in our survey data, so we chose point-like sources based on concentration indices ($\Delta i_{4-8}$): i.e., the difference between four- and eight-pixel diameter aperture-corrected $i$-band magnitudes (the median NGVS $i$-band seeing is $0\farcs54$; \citealt{Fer12}). We selected objects with $-0.05 \leq \Delta i_{4-8} \leq 0.6$, a range slightly wider than in \citet{Dur14}, to allow for the existence of larger GCs in UDGs \citep{vanD18,vanD19}. Among point-like sources, we then selected GCs using their $u^*g'i'$ colors, with the GC selection region in $u^*g'i'$ color space determined from spectroscopically confirmed GCs in M87 (see \citealt{Mun14} and especially \citealt{Lim17} for details). 

We must make two assumptions to compute the total number of GCs associated with a galaxy: (1) the effective radius of the GC system; and (2) the shape of the GC luminosity function (GCLF). We take the effective radius of the GC system to be $R_{e,GCS} = 1.5 R_{e,gal}$, where the effective radius of the galaxy is derived from NGVS imaging. 
In practice, this assumption means that half of the GCs associated with a given galaxy are found within 1.5 times the galaxy effective radius (e.g., \citealp{Lim18}). Under this assumption, the number of GCs within this aperture was counted (after applying a background correction) and then doubled to arrive at an estimate of the total number of observable GCs associated with each galaxy. In some cases, no GC candidate is detected within $1.5 R_{e,gal}$, so we expanded the aperture (up to $5R_e$) to have at least one GC candidate within the aperture. To correct for the spatial coverage of these enlarged apertures, we assume the GC number density profile to be a \citet {Ser63} function with $R_{e,GCS} = 1.5 R_{e,gal}$ and S\'ersic index $n=1$. When this was done, it has been noted in Tables~\ref{tbl:gcsn} and \ref{tbl:gcsn2}. Although some other studies have fitted GC spatial distributions directly to estimate the total number of GCs, the technique used in this paper is able to provide a homogeneous estimate of GC numbers in a diverse set of galaxies, from those with large GC populations to those containing few or no GCs.  Note that the total number of GCs in Coma UDGs estimated with this method are consistent with those measured directly from GC profile fitting \citep{Lim18}. We discuss the shape of the GCLF in Section~\ref{sec:gclf}.

In practice, our GC selection includes contaminants such as foreground stars, background galaxies, intracluster GCs, and GCs belonging to neighboring galaxies. We therefore estimated a local background by choosing eight local background regions, with each box-shaped region having an area of three square arcminutes. The mean and standard deviation in the number density of GC candidates in these background regions was then used to estimate the total numbers of GCs and their uncertainties in each UDG.

\subsubsection{GC Luminosity Function} \label{sec:gclf}

\begin{figure} 
\epsscale{1.2}
\plotone{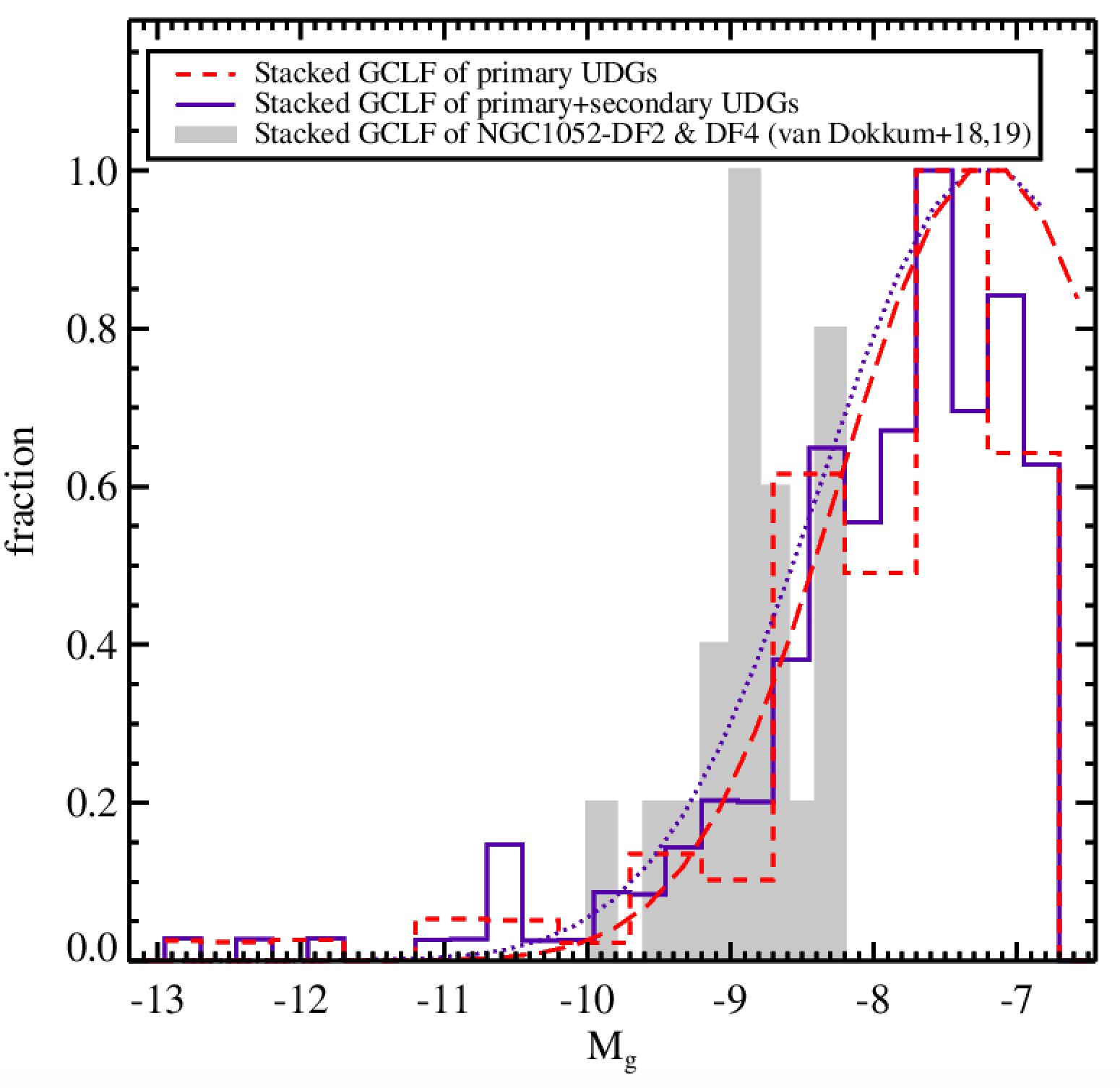}
\caption{Composite globular cluster luminosity function for Virgo UDGs. The dashed and solid histograms show GC candidates in the {\tt primary} and the {\tt primary+secondary} sample, respectively. Both samples are background subtracted. Light grey filled histograms represent stacked GC luminosity functions for NGC1052-DF2 and NGC1052-DF4. Both NGC~1052 group galaxies appear to have anomalously large and bright GC populations \citep{vanD18,vanD19}. Gaussian functions, fitted to the distributions of the primary and combined samples, are shown as the purple dotted and red long-dashed curves, respectively. For these curves, we fix the Gaussian mean to be $\mu_{g,TO}=-7.2$~mag, and find the best-fit $\sigma_g=1.0$~mag\label{gclf}} 
\end{figure}

GC luminosity functions are typically well described by a Gaussian function. The mean magnitude of the GCLF has a nearly universal value ($\mu_g=-7.2$~mag) with little dependence on host galaxy properties, while sigma has a correlation with the host galaxy luminosity \citep{Jor07,Vil10}. 
Recent studies of two UDGs (NGC1052-DF2 \& -DF4; \citealp{vanD18,vanD19}), however, suggest that these galaxies have GCLFs with mean magnitudes about $1.5$ magnitude brighter than a standard GCLF (although there has been some debate about their distances; \citealt{Tru19}). It would be scientifically interesting if the Virgo UDGs have a significantly different GCLF from other galaxies, but it would also introduce an additional source of uncertainty when we try to estimate the total number of GCs.

To test whether the form of the GCLF in Virgo UDGs is, on the whole, different from a standard GCLF, we have constructed a ``stacked'' GC luminosity function using the {\tt primary} and {\tt primary+secondary} samples. Figure \ref{gclf} shows the composite, background-subtracted GC luminosity functions. These luminosity functions are well fit with Gaussian functions down to our selection limit, and their mean magnitudes are consistent with the universal GC luminosity function. Overall, we do not find any significant excess at the peak luminosity of NGC1052-DF2 and -DF4's GCs. Additionally, although the numbers are small, we do not find any individual GC systems with an obvious LF peak around $M_g\approx-9$~mag. This result suggests that the form of the GCLF in Virgo UDGs is likely similar to those in other low-mass early-type galaxies \citep{Jor07,Mil07,Vil10}. 

Ultimately, we adopted a Gaussian GC luminosity function with parameters $\mu_g=-7.2$~mag and $\sigma_g=1.0$~mag, the latter of which was estimated from the stacked {\tt primary} sample GCLF with $\mu_g$ fixed. Our GC selection has a limiting magnitude of $g'\simeq24.5$~mag (at which we are 95\% complete), which is slightly deeper than the turn-over magnitude of GCLF at Virgo distance ($\mu_{g,TO}=23.9$~mag), so we should detect $\sim\!\!73\%$ of the GCs in a Gaussian distribution. To estimate the full number, we extrapolate the remainder of the GCLF using our assumed Gaussian LF. We note that $\sigma_g=1.0$~mag is consistent with what is seen in low-mass dwarfs \citep[$0.5\lesssim\sigma_g\lesssim1.3$~mag;][]{Jor07,Mil07}.

\subsubsection{GC Specific Frequencies}

\begin{figure*}[th] 
\epsscale{1.2}
\plotone{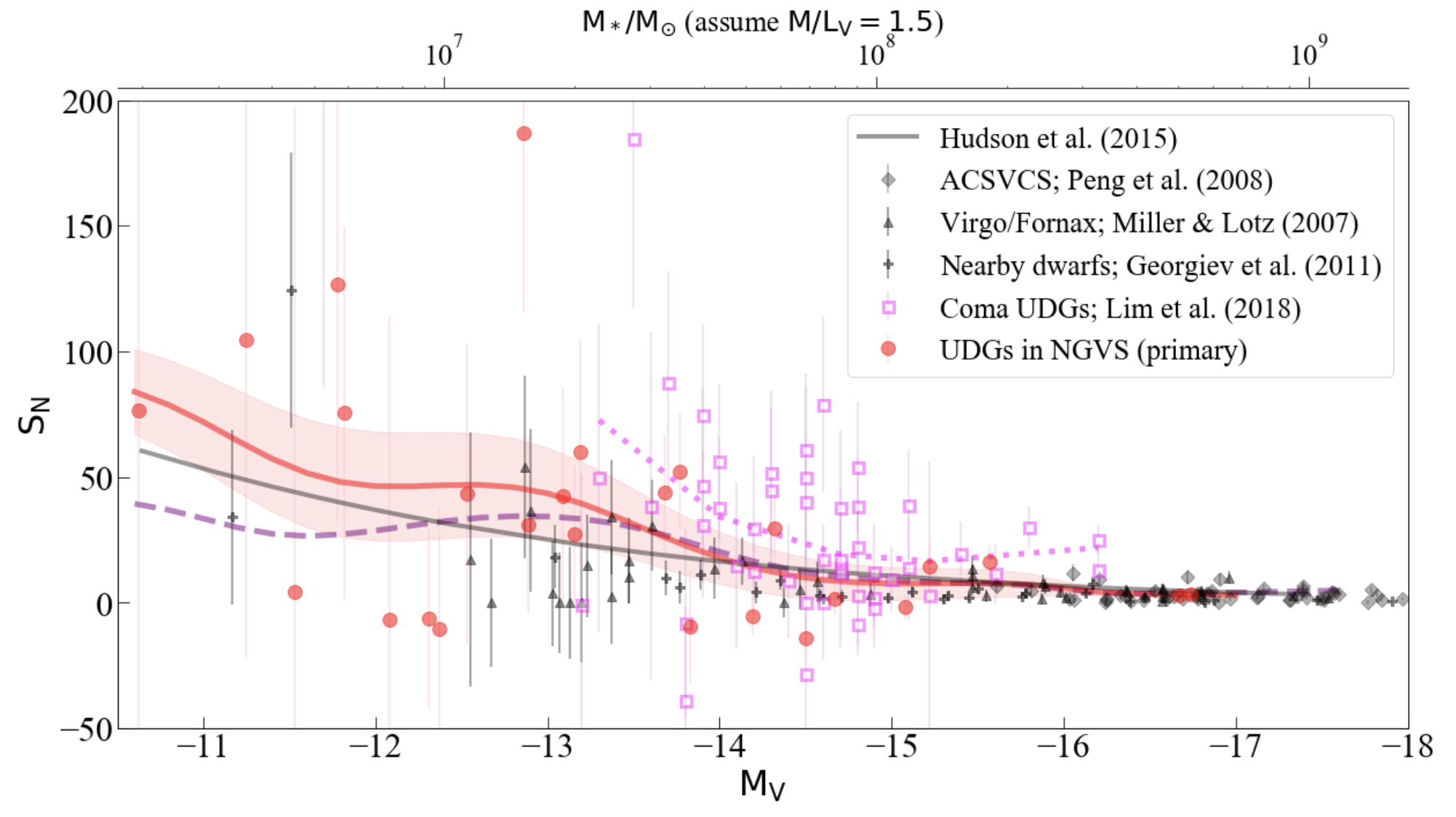}
\caption{Globular cluster specific frequencies, $S_N$, for Virgo cluster UDGs and comparison galaxies plotted as a function of absolute $V$-band magnitude. On the top axis, we show the corresponding stellar mass assuming $M/L_V=1.5$. 
The red filled circles show the {\tt primary} UDG sample. The red solid line and purple dashed line display the mean values of specific frequencies for the {\tt primary} sample and combined {\tt primary+secondary} sample, respectively. The red shaded region shows the uncertainty in the running mean for the {\tt primary} sample. The magenta squares and dotted magenta line show individual and mean $S_N$ values for Coma UDGs from \citet{Lim18}.
The black triangles, diamonds, and crosses show ``normal" early-type galaxies from \citet{Pen08}, \citet{Mil07}, and \citet{Geo10}, respectively. The solid black curve shows the predicted trend for $S_N$ assuming that the number of GCs scales with the host galaxy's inferred halo mass following \citet{Har17}, which assumes the stellar-to-halo mass relation (SHMR) of \citet{Hud15}. (Note that $S_N$ can formally be negative due to background subtraction.)\label{mvsn}}
\end{figure*}

With the total number of GCs in hand, we can then compute the GC specific frequency, $S_N$ \citep{Har81}. To estimate the  $V$-band magnitude of the galaxies, we use our $g'$-band magnitudes with an assumed $(g'-V)=0.1$~mag color for all galaxies. GC specific frequencies for the {\tt primary} sample of UDGs are compiled in Table~\ref{tbl:gcsn}, and for the {\tt secondary} sample of UDGs are compiled in Table~\ref{tbl:gcsn2}.

Figure~\ref{mvsn} compares the specific frequencies for Virgo UDGs in the {\tt primary} sample (red symbols) to those found in other types of galaxies and environments (i.e., Fornax, Coma and nearby dwarfs). In this plot, specific frequencies for high- and intermediate-luminosity early-type galaxies from the ACS Virgo Cluster Survey 
\citep[ACSVCS;][]{Cot04} are shown as open triangles \citep{Pen08}, with lower mass early-type dwarfs from \citet{Mil07} and \citet{Geo10}, and Coma cluster UDGs from \citet{Lim18}. Although the uncertainties in $S_N$ at such low stellar masses are large for any one galaxy,  the smoothed running mean (red line) does show a steady rise toward low masses, with $\langle S_N\rangle\sim70$ at $M_V=-11$~mag ($M_\star\sim3\times10^6 M_\odot$). We also show the running mean for the combined {\tt primary+secondary} sample (purple dashed line). 

The Virgo cluster UDGs on average have higher $S_N$ than classical dwarf galaxies in Virgo and Fornax, but lower than Coma cluster UDGs at comparable luminosities. The combined sample has a lower mean $S_N$ at low masses, suggesting that the {\tt secondary} sample galaxies are more like classical dwarfs. Fornax cluster UDGs have shown a similar trend \citep{Pro19b}. \citet{Lim18} also found that Coma cluster UDGs have systematically higher $S_N$ than classical dwarf counterparts at fixed stellar mass. In all cases, however, the scatter is large, with some UDGs having no GCs, and some having extremely high $S_N$. A direct comparison between the Virgo and Coma UDG populations is challenging given that many of the Virgo UDGs are fainter than those of Coma UDGs, and the extreme faintness of the Virgo systems means that the measurement of their effective radii is more difficult; as a result, the specific frequencies for Virgo UDGs have larger uncertainties than their Coma counterparts.
Previous observational and theoretical studies \citep{Pen08,Mis16,Lim18} have shown that low mass galaxies in denser environments can have higher $S_N$. It is possible that similar processes may explain the difference in $S_N$ between the Coma UDGs and the ones in the Virgo and Fornax clusters.

There is increasing evidence that the number of GCs (or the total mass of the GC system) correlates better with total galaxy halo mass than with stellar mass \citep[e.g.,][]{Bla97,Pen08,Har17}, although the reason why this might true is still under debate \citep{Boy17,ElB19,Cho19}. Very few galaxies have both GC numbers and directly measured halo masses. What is typically done is to assume a stellar-to-halo mass relation (SHMR), and then estimate the total mass fraction of GCs. We show the implications of this assumed relation with the gray line in Figure~\ref{mvsn}. This curve is calculated in the same way as in the study of \citet{Har17}, using the SHMR from \citet{Hud15} evolved to redshift zero. We then extrapolate the SHMR to $10^6 M_\odot$ and assume $\eta = 2.9\times10^{-5}$, where $\eta$ is the number of GCs per solar mass of halo mass \citep{Bla97}.

We note, however, that the SHMR of \citet{Hud15} is not calibrated below $\sim\!\!2\times10^{10} M_\odot$ for quenched galaxies, so our use of this relation is an extrapolation of over four orders of magnitude. Additionally, this and most other SMHRs are for centrals, while most of the data used to calibrate $\eta$ is from satellites. Assuming different SHMRs may be informative, like those for satellites in Virgo \citep{Gro15}, but we should then re-estimate $\eta$ using the appropriate data. We leave a more involved discussion of this subject for a future paper, and simply note that the mean trend of $S_N$ with $M_V$ for UDGs is generally consistent with $\eta = 2.9\times10^{-5}$ down to low mass when the \citet{Hud15} SMHR is extrapolated to low masses. However, given the scatter in $S_N$ for the UDGs, one should be careful in trying to invert this relation and estimate individual UDG halo masses from GC numbers or masses.

\subsubsection{GC Color Distributions}

\begin{figure} 
\epsscale{1.2}
\plotone{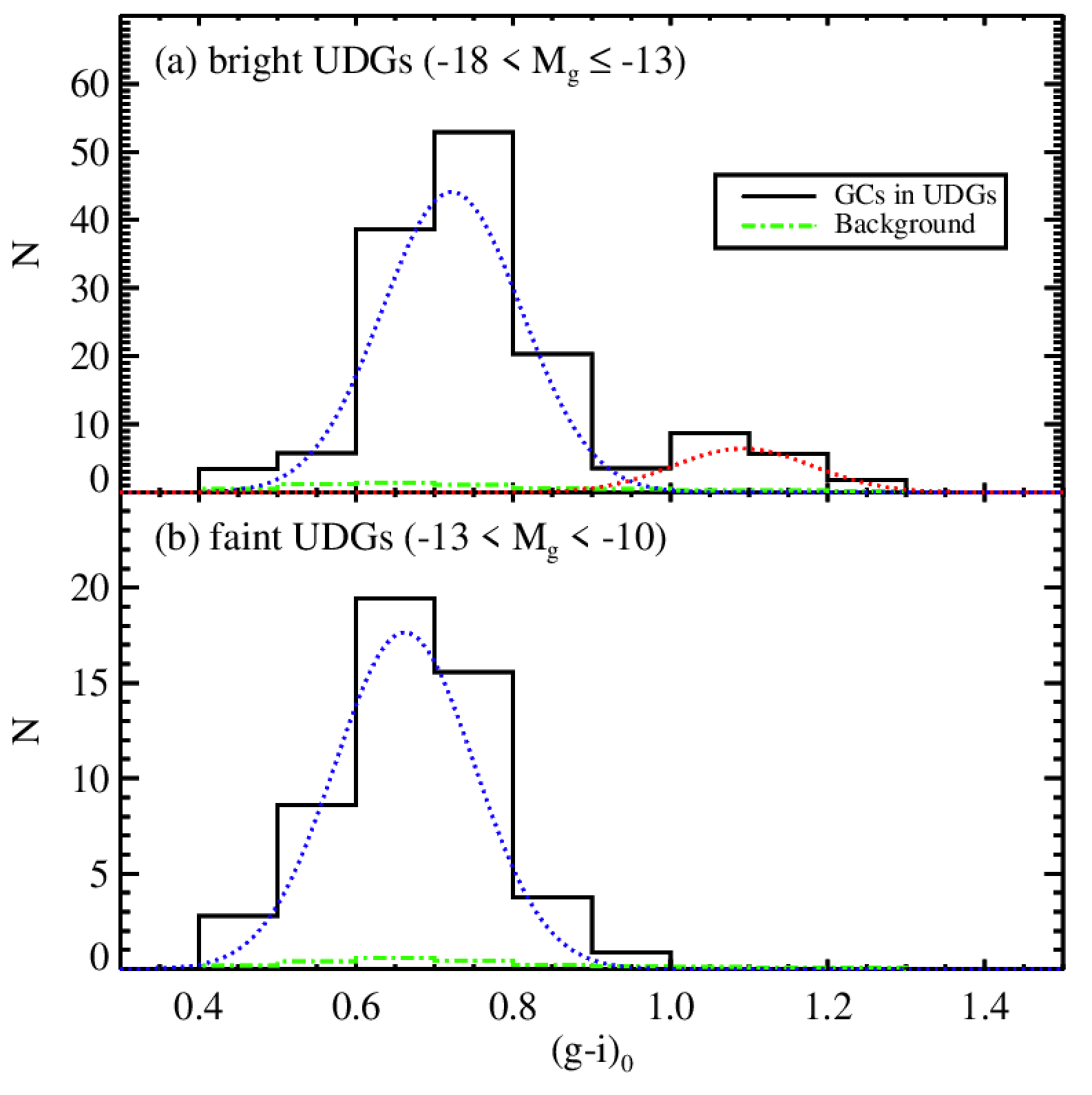}
\caption{Composite globular cluster $(g-i)_0$ color distribution for Virgo UDGs. The solid and dashed-dot histograms show GC candidates in the {\tt primary+secondary} sample and GC candidates in the adjacent backgrounds, respectively. The solid histograms are background subtracted. {\it Panel (a)} and {\it Panel (b)} show the bright UDGs ($-18 < M_g \leq -13$) and faint UDGs ($-13 < M_g \leq -10$), respectively. Two Gaussian functions, fitted simultaneously to the background-subtracted distribution, are shown as the blue and red dotted curves, but {\it Panel (b)} only has a blue curve.
\label{gccol}} 
\end{figure}

The colors of GCs, as rough proxies for their metallicities or ages, have been characterized across a wide range of galaxies \citep[e.g.,][]{Lar01,Pen06}, and can provide insights into the formation and evolution of their host galaxies. GC metallicities, represented in a crude way by their broadband colors under the assumption of old ages, have been observed to have a wide spread in massive galaxies, with both metal-poor (blue, possibly accreted) and metal-rich (red, possibly formed {\it in situ}) populations. The mean colors and relative fractions of both of these populations have been shown to be correlated with the stellar mass of the host, although the slope is much steeper for the metal-rich GCs, especially when GC color gradients are taken into account \citep{Liu11}. The color distributions of GCs in UDGs, both their mean colors and the relative fractions of blue and red GCs, have the potential to tell us about their chemical enrichment history, although the exact translation of colors to metallicity in old GCs is still a subject of debate \citep[e.g.,][]{Pen06,Yoo06,Bla10,Villaume19,Fah20}. 

Unfortunately, due to the  small number of GCs typically associated with any individual UDG --- and the lack of multi-color imaging in a number of previous surveys --- few UDGs have had their GC color distributions studied in any detail, with DF17 \citep{Bea16b} and DF44 \citep{vanD17} in the Coma Cluster being the exceptions. In DF17, the mean GC color of $\langle(g-z)\rangle = 1.04$, roughly equivalent to $\langle(g'-i')\rangle \approx 0.79$, is a bit redder than expected for a galaxy of its stellar mass, and similar to those of GCs in galaxies with virial masses of $\sim\!10^{11}M_{\odot}$.

Although some of the Virgo UDGs studied here have large GC populations relative to their stellar luminosity, their low absolute numbers make difficult a case-by-case study of their color distributions. We have therefore constructed combined GC color distributions using a sample of 36 UDGs from the combined {\tt primary+secondary} sample that have GC detections within $1.5R_e$. The only galaxy with GC candidates that is excluded is NGVSUDG-A09 (VCC~1249), because its close proximity to M49 makes its GC color distribution extremely uncertain. We further divide this sample into two bins of luminosity ($-18< M_g<-13$ and $-13\leq M_g<-10$, the dividing point roughly corresponding to $M_\star\sim2\times10^7 M_\odot$), with 19 and 17 galaxies in the ``bright'' and ``faint'' samples, respectively. Figure \ref{gccol} presents the resulting composite background-subtracted $(g'-i')_0$ GC color distributions. 

The ``bright'' UDG sample shows an apparent bimodality in GC colors similar to what is seen in more massive galaxies. We fitted this distribution with two Gaussian functions using Gaussian Mixture Modeling (GMM) code \citep{Mur10}. GMM provides best-fit Gaussian parameters as well as a D value that indicates the separation of two peaks relative to their width. Fitting a pair of Gaussians is statistically justified when $D>2$, and the D value for our composite GC color distribution is $3.9\pm1.2$. The two Gaussian peaks are located at colors of $(g'-i')_0=0.72\pm0.01$~mag and $(g-i)_0=1.10\pm0.01$~mag. In terms of total numbers, we find 87\% and 13\% of the GCs belonging to the blue and red populations, respectively. The blue peak is consistent with the peak of the GC color distribution of the Fornax UDGs \citep{Pro19b}. The ``faint'' UDG sample has only a single peak of blue GCs whose mean color is $(g'-i')_0 = 0.66\pm0.01$~mag.

There are two interesting results here. First, the data suggest the existence of a significant (if small) population of red GCs in the ``bright'' UDG sample. Upon closer inspection, a majority of the red GCs are in the brightest UDGs in the sample, like NGVSUDG-05, -09, -26, and -A10. A couple of these (NGVSUDG-09 and -A10) show disturbed isophotes or shells indicating a possible interaction or post-merger state. 

Second, we can compare the mean colors of the different populations to each other and to those seen in normal early-type galaxies. Comparing the mean colors of the blue peaks shows a clear difference, where the fainter galaxies have a blue GC population that is bluer by $\Delta (g'-i')_0 = 0.06$~mag. To compare with the relations in \citet{Liu11}, we transform to the HST/ACS filter system using GCs that are in both the NGVS and ACS Virgo  observations:
\begin{equation}
(g'-z')_{0,ACS} = 1.65\times(g'-i')_{0,NGVS} - 0.27
\end{equation}
For the ``bright'' sample. the blue and red peaks thus have mean colors of $(g'-z')_0=0.92\pm0.02$~mag and $(g'-z')_0=1.54\pm0.16$~mag, respectively, which we can compare to Figure~6 in \citet{Liu11}. Despite having a low red GC fraction that is consistent with what we would expect for galaxies at this mass \citep{Pen08}, we find that both the blue and red GCs in the ``bright'' UDG sample have mean colors that are much redder than expected for the stellar mass of their hosts. These UDGs are $100\times$ less massive than the ACSVCS galaxies that host GCs with similar colors. None are obviously near massive galaxies whose more metal-rich GC systems may be sources of contamination. The ``faint'' sample has a single peak at $(g'-z')_0=0.82\pm0.02$~mag. As a contrast to the ``bright'' sample, it has very blue GCs, with a mean color bluer than those in the least massive ACSVCS galaxies, and consistent with being an extension of the previously established relationship between blue GC mean color and galaxy luminosity. 

We have inspected the red GC candidates in the individual UDGs. Typically, just one or two extreme objects, with $(g-i)_0\gtrsim1.0$, are found in any individual UDG, so this population is not free from uncertainties due to small number statistics and imperfect background subtraction. Moreover, a number of red GCs are located far from their galaxy centers (i.e., 11 of the 15 red GCs in {\tt primary} UDGs are found beyond the effective radius of their host galaxy). We suspect some of these objects may be due to residual contamination by background objects. Radial velocity confirmation of membership and spectroscopic age and metallicities for these objects will be needed to establish their true nature.

\subsubsection{Nuclear Star Clusters}

\citet{San19} studied the fraction of nucleated galaxies in the Virgo core region, and showed that it varies from $f_{nuc}\approx0$ to $f_{nuc}\approx90\%$ depending on the stellar mass of the host galaxy. We have examined the evidence for stellar nuclei in the NGVS isophotal models and through visual inspection, and find that 3-4 of the 26 UDGs in our {\tt primary sample} appear to be nucleated\footnote{These are NGVSUDG-01, NGVSUDG-06, NGVSUDG-26 and, possibly, NGVSUDG-04. Likewise, 3-5 galaxies in our {\tt secondary} sample may also be nucleated: i.e., NGVSUDG-A08, NGVSUDG-A11, NGVSUDG-A15 and, possibly, NGVSUDG-A03 and NGVSUDG-A10}. Throughout the entire cluster, the overall UDG nucleation fraction is therefore $f_{nuc,UDG}= (6-9)/44 \simeq 14-20\%$. The nucleation fraction in the core is similar: i.e., 2 of the 10 UDGs (20\%) belonging to the combined {\tt primary} and {\tt secondary} sample appear to be nucleated. For comparison, the nucleation fraction of ``normal" galaxies, with luminosities similar to the UDGs, ranges from $20-60\%$ \citep{San19}. Thus, the UDGs may have a slightly lower nucleation fraction than other Virgo galaxies, consistent with our recent findings in Coma \citep{Lim18}.

\section{Discussion}
\label{discussion}

\subsection{The Uniqueness of UDGs as a Population}
\label{unique}

Based on the residuals from the mean scaling relations observed in the core region (see Figures~\ref{dSBe}, \ref{dmSBe} and \ref{dRe}), the 10 UDGs in the central $\sim4$ deg$^{2}$ ({\tt primary} and {\tt secondary} samples combined)  seem to be marginally distinct, slightly separated from the population of $\sim$400 ``normal" galaxies. However, a different picture emerges when one considers the full sample of $\sim$3700 galaxies that are distributed throughout the cluster. With an order-of-magnitude larger sample size, the gaps in effective radius and surface brightness are no longer apparent, and the UDG candidates (26 or 44 galaxies, depending on which sample is used) seem to occupy the tails of Gaussian-like distributions in structural parameters. While it is entirely possible that scaling relations of normal and diffuse galaxies depend on environment --- and perhaps behave differently for the two populations --- it is also possible that the gaps seen in the core region are an artifact of the smaller sample. Our provisional conclusion is that, when one considers the cluster in its entirety, UDGs are simply galaxies that occupy the LSB tail of the full population. Of course, this interpretation does not rule out the possibility that the galaxies that populate this LSB tail do so because they have been prone to physical processes, such as tidal heating and disruption (e.g. \citealp{Car19}), that may give rise to at least some UDGs. 

\subsection{The Spatial Distribution of UDGs as a Clue to their Formation}
\label{spatial}

Our study differs from most previous UDG surveys in that we target a single environment with high and nearly uniform photometric and spatial completeness: i.e., roughly speaking, the NGVS reaches detection limits of $g\sim25.9$~mag and $\mu_g \sim 29$ mag~arcsec$^{-2}$, over the entire $\sim\!\!100$ deg$^2$ region contained within the cluster virial radius. Thus, it is possible with the NGVS to explore the spatial distribution of UDGs within the cluster, compare it to that of normal galaxies, and use this information to critically assess formation scenarios.

Figure~\ref{cumudg} shows one of the principal findings of this paper: {\it the Virgo UDG candidates are more spatially concentrated on the central region than other cluster members}. This is true for both the {\tt primary} and the combined {\tt primary} and {\tt secondary} samples. This finding is noteworthy because previous studies --- often relying on incomplete or heterogeneous data --- have reached conflicting conclusions on whether or not UDGs favor the dense central regions of rich clusters (e.g. \citealp{vanB16,Man18}). It is worth bearing in mind that the UDG candidates in Virgo extend to significantly lower luminosities and surface brightness levels than those uncovered in previous surveys (e.g., fully half of the Virgo UDGs are fainter than $\langle\mu\rangle_{e}$ = 26 mag arcsec$^{-2}$). Deeper imaging and/or expanded spatial coverage of other clusters, with a consistent definition of the UDG class, will be required to know if Virgo is unique in this sense. 

Unlike previous UDG studies, the selection criteria used in this study rely on the {\it empirical trends} between luminosity and structural parameters ($R_e$, $\Sigma_e$, $\langle\Sigma\rangle_e$) defined by a nearly complete sample of cluster members in the core region, as well as the {\it observed scatter} about these mean relations. Interestingly, we find the core to be overabundant in UDG candidates relative to the rest of the cluster. For example, 10/44 (23\%) of the galaxies in our combined {\tt primary} and {\tt secondary} samples, are found in the central $\sim 4$~deg$^2$. Although we cannot rule out the possibility that some of these candidates are being seen in projection against the cluster core, there are reasons to believe the overall enhancement is real. While it is difficult to assess the importance of projection effects without an {\it a priori} knowledge of the three-dimensional (volume) density distribution, we can nevertheless test the possibility that the observed excess is due to random chance. To illustrate, a total of 2101 cluster galaxies have luminosities in the range defined by the combined {\tt primary} and {\tt secondary} samples. In random selections of these galaxies, carried out 5000 times, a total of 10 or more galaxies fall in the central 4 deg$^ 2$ in only $1.5\%$ of the cases. This suggests that the observed central enhancement is real, and not purely the result of projection effects. 

This central concentration of UDGs in the Virgo core region is puzzling. \citet{Ron17} investigated a possible dwarf origin for UDGs using the Millennium II cosmological simulation and Phoenix simulations of rich clusters. Comparing to a variety of observations, they concluded that a dwarf origin for UDGs is feasible since the predicted objects match the observations in a number of cases, including their spatial distribution and apparent absence in the central regions of clusters like Coma. Furthermore, tidal disruption modeling within the hydrodynamical simulations IllustrisTNG showed that UDGs might have a dual origin: a sub-population of dwarf-like halos with late infall, in agreement with \citet{Ron17}, and another sub-population resulting from tidal disruption of more massive galaxies with remnants consistent with the properties of UDGs \citep{Sal20}. It would be worthwhile to revisit these theoretical results in light of our Virgo observations. For example, \citet{Sal20} predict that a tidal origin might be confirmed by UDGs showing a combination of low velocity dispersion and an enhanced stellar metallicity.  Encouragingly, the distribution of UDGs within these Virgo-like clusters in IllustrisTNG is also peaked towards the cluster centers, in good agreement with our findings in Virgo. 

We note that the apparent lack of tidal features in some UDG samples may not rule out tidal disruption as an important formation process. For instance, our previous study of the kinematics of GCs in VLSB-D shows clear evidence for on-going disruption \citep{Tol18} that could have been missed from shallower surveys or lack of kinematics information for the GCs (and see \S\ref{morphology} and Appendix~B for some specific evidence for this UDG and others). It is important to bear in mind that the mere {\it detection} of UDGs can be challenging given their low surface brightness, and faint tidal features even more so.

\subsection{Clues from Morphologies}
\label{morphology}

We now pause to consider the question of UDG morphologies, and what clues they may hold for formation models. 
Thumbnail images for our sample of UDGs can be found in Figure~\ref{udgthumb_primary}  and \ref{udgthumb_secondary}. Although the majority of these galaxies are, by their nature, faint and diffuse objects, a careful inspection of the NGVS images, combined with an analysis of the best-fit models from ELLIPSE/BMODEL and/or GALFIT (see \S\ref{data}), offers some clues to the origin of UDGs, and their (non)uniformity as a class. For instance, the 44 galaxies belonging to our {\tt primary} and {\tt secondary} samples exhibit a wide range in axial ratio: i.e., a number of objects are highly flattened but many others have a nearly circular appearance. This na\"ively suggests that a tidal origin, which may be a viable explanation for some UDGs, is unlikely to account for all members of this class. 

Nevertheless, a tidal origin seems likely, if not certain, for some UDGs. As noted in the Appendix, we see evidence for tidal streams associated with at least four galaxies: NGVSUDG-01 (VCC197), NGVSUDG-A07 (VCC987), NGVSUDG-A08 and NGVSUDG-A09 (VCC1249). Within this small subsample, there is one object that belongs to an infalling group located on the cluster periphery (VCC197; \citealt{Pau13}), two galaxies that are deep within the cluster core (VCC987 and NGVSUDG-A08), and one low-mass, star-forming galaxy (VCC1249; \citealt{Arr12}) that is tidally interacting with M49, the brightest member of the cluster.

A number of other UDGs clearly have disturbed morphologies --- such as twisted or irregular isophotes, shells and ripples --- that are indicative of post-merger, or post-interaction, galaxies: i.e., NGVSUDG-02 (VCC360), NGVSUDG-09 (VCC1017), NGVSUDG-10 (VCC1052) and NGVSUDG-A10 (VCC1448). Additionally, a handful of UDGs --- most notably NGVSUDG-08 and NGVSUDG-A14 --- may be members of LSB pairs, while at least one object --- NGVSUDG-A11 --- shows clear evidence for a faint spiral pattern at large radius, despite its previous classification as a dE0,N galaxy \citep{Bin85}.

In short, their morphologies demonstrate that at least some of the objects found in the LSB tail of the ``normal" galaxy population probably owe their diffuse nature to physical processes --- such as tidal interactions or low-mass mergers --- that are at play within the cluster environment. Likewise, the diversity in their morphologies provides {\it prima facie} evidence that no single process has given rise to all objects within the UDG class. It will be valuable to investigate UDG morphologies more closely for a subset of the Virgo objects, ideally with deep, high-resolution images that can be used to map ultra-LSB features using individual RGB stars.

\section{Summary}
\label{summary}

As part of the Next Generation Virgo Cluster Survey \citep{Fer12}, we have identified and characterized UDGs in the nearby Virgo Cluster. Employing a new, quantitative definition for UDGs based on the structural parameters of galaxies in the Virgo Cluster core (i.e., luminosity, effective radius, effective surface brightness and mean effective surface brightness; \citealt{Cot20}), we have identified candidate UDGs throughout the cluster, from the core to the periphery. In our analysis, we define two UDGs samples: (1) a {\tt primary} sample of 26 candidates selected as LSB outliers in each of three scaling relations, which ensures {\it high purity}; and (2) a combined {\tt primary} and {\tt secondary} sample of 44 UDGs which was assembled to ensure {\it high completeness}. Roughly half of these objects (21/44) are previously-cataloged galaxies, including eight galaxies previously identified as dwarfs of very large size and low surface brightness by \citet{Bin85}. 

Our principal conclusions are:

\begin{enumerate}
\item[{$\bullet$}] In a 4 deg$^{2}$ region in the Virgo core, which was used to establish our UDG selection criteria, we find 10 UDG candidates among a sample of 404 galaxies (an occurrence rate of $\sim$2.5\%). These candidates appear marginally distinct in their structural properties: i.e., separated by small gaps in effective radius and surface brightness from the population of ``normal" galaxies. However, when one considers the full sample of 3689 member galaxies distributed throughout the cluster, this separation vanishes. 

\item[{$\bullet$}] We compare the spatial distribution of our UDG candidates to ``normal" Virgo galaxies, and find the UDGs to be more centrally concentrated than the latter population, contrary to some findings in other clusters (e.g., \citealp{vanB16,Man18}). A significant number of UDGs reside in the core region, including some of the faintest candidates. Using the combined sample of 44 {\tt primary} and {\tt secondary} UDGs, 10 objects, or 23\% of the entire UDG population, are found in the core region (which represents less than 4\% of the cluster by area). Although we cannot rule out the possibility that some of these objects are seen in projection against the cluster core, the central enhancement is likely real and may be related to strong tidal forces in this region, or perhaps to the earlier infall expected of objects in this region.

\item[{$\bullet$}] Many of the UDG candidates in Virgo are exceptionally faint, and they expand the parameter space known to be occupied by UDGs. The faintest candidates have mean effective surface brightnesses of $\langle\mu\rangle_e \sim$ 29 mag arcsec$^{-2}$ in the $g$-band. Previous imaging surveys targeting UDGs in other environments have typically been limited to candidates brighter than $\langle\mu\rangle_e =$ 27.5 mag arcsec$^{-2}$ in the $g$-band. More than half of our Virgo UDG candidates are fainter than this.

\item[{$\bullet$}] We have carried out a first characterization of the GC systems of these galaxies. Although a direct comparison between the Virgo UDGs and those in other environments is complicated by the fact that the samples differ in luminosity and surface brightness, we find the Virgo UDGs to have GC specific frequencies that are slightly lower than those in Coma UDGs at comparable luminosities, yet somewhat elevated compared to ``normal" early-type dwarf galaxies. Consistent with recent findings in the Coma Cluster, the Virgo UDGs appear to show a wide range in their GC content. The mean $S_N$ of Virgo UDGs increases with decreasing stellar mass, roughly consistent with the expectation from a constant scaling between $N_{GC}$ and halo mass.

\item[{$\bullet$}] The GCs in these UDGs are predominantly blue. UDGs fainter than $M_g=-13$ have entirely blue, metal-poor GC populations. UDGs brighter than $M_g=-13$ have, $\sim\!13\%$ of their clusters in a red, metal-rich population. Moreover, the mean colors of both the blue and red GCs in the bright sample have colors that are significantly redder than in galaxies of comparable luminosity. The mean color of the blue GCs in the faint sample are consistent with an extrapolation of known scaling relations. The number of red GC candidates is small, and spectroscopy will be needed to confirm membership and establish their true nature.

\item[$\bullet$] In terms of morphology, there is clear diversity within the UDG class, with some objects showing evidence of a tidal origin whiles others appear to be post-merger or post-interaction systems. This suggests that no single process has given rise to all objects within the UDG class.

\item[$\bullet$] Weighing the available evidence --- and especially the apparent continuities in the size and surface brightness distributions (at fixed luminosity) when the cluster is considered in its entirety --- we suggest that UDGs may simply be those systems that occupy the extended tails of the galaxy size and surface brightness distributions. The physical mechanisms that shape the low (and high) surface brightness tails of the galaxy distributions remain interesting topics for future study.

\end{enumerate}

Some obvious extensions of this work present themselves. Radial velocity measurements for GCs in these UDGs will make it possible to measure dynamical masses and dark matter content. Deep imaging from the Hubble Space Telescope (and/or future space-based imaging telescopes) will allow the detection of individual RGB stars in these galaxies, and allow some key  parameters to be measured, such as distance, chemical abundance, mean age and surface density profiles. Such images would additionally enable a search for LSB tidal features, and perhaps provide insight into the role of tidal forces in the formation of these extreme galaxies. 

\acknowledgments
This research was supported by Basic Science Research Program through the National Research Foundation of Korea (NRF) funded by the Ministry of Education (NRF-2018R1A6A3A03011821).
EWP acknowledges support from the National Natural Science Foundation of China (11573002), and from the Strategic Priority Research Program, ``The Emergence of Cosmological Structures,'' of the Chinese Academy of Sciences (XDB09000105). C.L. acknowledges support from the National Natural Science Foundation of China (NSFC, Grant No. 11673017, 11833005, 11933003, 11203017)

\appendix
\section{Sample Selection and Additional UDGs Candidates}
\label{sec:appendix1}

As discussed in \S\ref{selection}, Virgo UDGs are selected using the observed ($g'$-band) scaling relations of ``normal" Virgo Cluster galaxies --- specifically the three relations involving effective radius ($L$-$R_e$), effective surface brightness ($L$-$\mu_e$) and mean effective surface brightness ($L$-$\langle\mu\rangle_e$). The 26 galaxies that deviate by 2.5$\sigma$, or more, toward low surface brightness in each of these three relations comprise our {\tt primary} sample. Thumbnail images for these 26 galaxies are shown in Figure~\ref{udgthumb_primary} and detailed notes are given below (\S\ref{notes:primary}).

\begin{figure*}
\epsscale{1.2}
\plotone{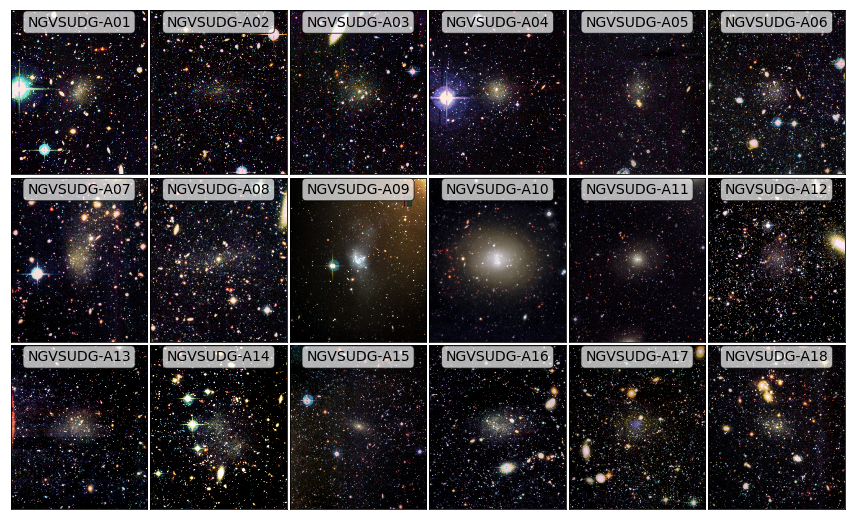}
\caption{Thumbnail images of 18 UDGs identified among our {\tt secondary} sample of LSB outliers. The field of view is $4\farcm4 \times 5\farcm3$. North is up and East is left. To enhance low surface brightness features, all images have been smoothed using a Gaussian kernel with $\sigma = 2.5$~pixels (${\rm FWHM} =1\farcs1$). The ID of each UDG is noted at the top of the image.  \label{udgthumb_secondary}}
\end{figure*}

\begin{figure*}
\epsscale{1.2}
\plotone{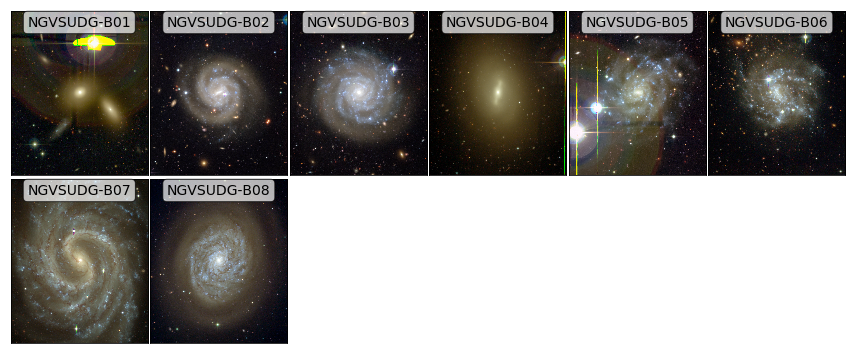}
\caption{Thumbnail images of eight bright, face-on, spiral galaxies identified among our {\tt secondary} sample of LSB outliers. The field of view is $4\farcm4 \times 5\farcm3$. North is up and East is left. To enhance low surface brightness features, all images have been smoothed using a Gaussian kernel with FWHM = 3 pixels ($0\farcs66$). The ID of each UDG is noted at the top of the image. \label{udgthumb_spiral}}
\end{figure*}

Although this {\tt primary} sample is expected to have high purity, it is likely that some bonafide LSB galaxies may be missed given the requirement that a galaxy deviates significantly in all three scaling relations. To ensure a more complete catalog of UDGs, we therefore define a {\tt secondary} sample by relaxing the selection criteria to include galaxies that deviate by $\ge$ 2.5$\sigma$ in two, or even one, of the relations. A total of 26 galaxies were selected as LSB outliers using these less restrictive criteria. As we show below, these galaxies fall into two classes: (1) 18 galaxies that appear to be genuine UDGs; and (2) eight bright, face-on spiral/S0 galaxies. These samples are shown in Figures~\ref{udgthumb_secondary} and \ref{udgthumb_spiral}, and notes on individual galaxies are presented in \S\ref{notes:secondary} and \ref{notes:spirals}, respectively.

\section{Notes on Individual Galaxies}
\label{sec:appendix2}

\subsection{Primary Sample}
\label{notes:primary}

\begin{itemize}

    \item {\tt NGVSUDG-01}. This large ($R_e \simeq 63\arcsec$) and low surface brightness galaxy ($\mu_e \simeq 28.9$ mag~arcsec$^{-2}$) was identified by \citet{Rea83}. It appears in \citet{Bin85} as VCC197 although it was not listed among their sample of dwarfs ``of very large size and low surface brightness." It is a highly elongated galaxy ($q \sim 0.3$ and $\theta \sim 65^{\circ}$ that is, in fact, embedded in a long, narrow filamentary tidal stream (\citealt{Pau13}; see also \citealt{Mar10}) within the infalling NGC4216 (VCC167; $131\pm4$\kms) group on the WNW edge of the cluster, located roughly 3.6$^{\circ}$ ($\sim1$~Mpc) from M87. The galaxy appears to contain a bright nucleus at its photocenter.  In addition to VCC197 itself, and the dominant spiral, VCC167, a number of other VCC galaxies are confirmed radial velocity members of this group (VCC187, VCC165 and VCC200) as well as several new likely group members identified in the NGVS (NGVSJ12:16:47.13+13:12:19.4, NGVSJ12:16:08.32+13:06:45.7).
    
    \item {\tt NGVSUDG-02.} This relatively bright UDG, VCC360, is located in the remote NW region of the cluster, 4\fdg1 (1.2~Mpc) from M87 and 52$^\prime$ (250~kpc) from M100 (NGC4321 = VCC596). It was discovered by \citet{Rea83} and cataloged by \citet{Bin85}, although not identified as an extreme LSB dwarf in their Table~XIV. It is a  peculiar galaxy, with a seemingly disturbed morphology: i.e., an {\tt X} structure in its core with loops/tails at large radii, reminiscent of a dwarf-dwarf post-merger system (e.g., \citealt{Zha20}). This object contains a few candidate GCs and a newly detected dwarf (NGVSJ12:19:16.39+15:23:46.3) is located 6$^\prime$ (29~kpc) to the~SW.
    
    \item {\tt NGVSUDG-03.} This previously uncatalogued object ($g' \simeq 18.4$~mag, $R_e \simeq 33$\arcsec and $\mu_e \simeq 28.3$ mag~arcsec$^{-2}$) is located $\sim$2\farcm5 (12~kpc) to the WNW of VCC674. It is one of several faint galaxies discovered in this region by the NGVS, including two very faint systems within 3$^\prime$ (NGVSJ12:24:16.46+13:51:24.7 and NGVSJ12:24:08.52+13:49:59.6). 
    A handful GC candidates may be associated with the galaxy.
    
    \item {\tt NGVSUDG-04.} This moderately bright ($g' \simeq 17.6$) but large ($R_e \simeq 33\arcsec$) and low surface brightness ($\mu_e \simeq 27.9$ mag~arcsec$^{-2}$) galaxy was first identified by \citet{Mih15} as VLSB-D.  It is located 1\fdg9 (550~kpc) from M87, offset in the NW direction, in a relatively isolated part of the cluster. 
    Several GC candidates are visible, along with a bright source close to the galaxy photocenter that may be a nuclear star cluster ($g' \simeq 20.7$~mag). \citet{Tol18} measured velocities for 12 GCs in this system with a mean velocity of $v_r = 1034\pm6$~\kms, a velocity dispersion of $\sigma = 16^{+6}_{-4}$~\kms, and a velocity gradient along the major axis. The elongated appearance and velocity gradient suggest that it may be in the process of being tidally stripped.
    
    \item {\tt NGVSUDG-05.} This galaxy, located 2\fdg5 from M87 in the SW direction, was first identified by \citet{Bin85} as VCC811 and listed as one of their large and low surface brightness dwarfs.  It is a relatively bright and extended galaxy ($g' \simeq17.0$~mag, $R_e \simeq 34\arcsec$ and $\mu_e \simeq 27.1$ mag~arcsec$^{-2}$) with a round appearance ($q \sim 0.95$). A faint dwarf from the NGVS (NGVSJ12:25:33.36+10:13:53.0) is found 1\farcm5 (7.2~kpc) to the SW, and roughly ten GC candidates seem to be associated with the galaxy. 
    
    \item {\tt NGVSUDG-06.} This is VCC927, a typical UDG with $g' \simeq18.2$~mag, $R_e \simeq 26\arcsec$ and $\mu_e \simeq 28.1$ mag arcsec$^{-2}$. It is moderately flattened, with $q\sim0.8$ and $\theta \sim 110^\circ$, and located in a crowded part of the cluster, offset by 1\fdg2 deg (350 kpc) from M87 in the NW direction. A few GC candidates may be associated with the galaxy, with an apparent nuclear star cluster ($g' \simeq 21.2$~mag) at its photocenter.
    
    \item {\tt NGVSUDG-07.} This dwarf was previously identified as {\tt N lsb10} by \citet{Cal06} who noted its  low surface brightness nature. This very faint object ($g \simeq 19.0$~mag, $R_e \simeq 28\arcsec$ and $\mu_e \simeq 28.9$ mag arcsec$^{-2}$) is located in the cluster core, close to the centroid of the Virgo dwarf population and 1\fdg4 (400~kpc) from M87 in the NW direction. One or two candidate GCs may be associated with the galaxy. It is a fairly elongated object, with $q\sim0.7$ and $\theta \sim 103^\circ$. \citet{Cal06} report a TRGB distance of $15.85\pm0.88$ Mpc, a mean metallicity of ${\rm [Fe/H]} = -1.6 \pm 0.2$ and an absolute visual magnitude of $M_V = -11.6$ (about 0.7 mag fainter than our estimate; see Table~\ref{tbl:gcsn}).
    
    \item {\tt NGVSUDG-08.} This faint object ($g' \simeq 18.9$~mag, $R_e \simeq 21\arcsec$ and $\mu_e \simeq 28.3$ mag arcsec$^{-2}$) is newly detected in the NGVS. There is a hint in the CFHT images that this system may, in fact, be a double LSB galaxy, although this cannot be established with certainty. It is a moderately flattened galaxy, with $q \sim 0.8$ and $\theta \sim 12^{\circ}$, and no perceptible GC system. The nearest VCC galaxy is VCC1004, a faint dwarf located 3\farcm7 (18~kpc) to the SE.
    
    \item {\tt NGVSUDG-09.} This is VCC1017, a LSB dwarf with $v_r = 32\pm33$ \kms~  reported by \citet{Zab93} who noted the presence of both absorption and emission lines in their optical spectrum. The galaxy was discovered by \citet{Rea83}, and catalogued by \citet{Bin85} who listed it as a (ImV-type) dwarf of very large size and low surface brightness.  It is one of the brighter galaxies in our UDG sample, with $g' \simeq14.5$~mag, $R_e \simeq  54\arcsec$ and $\mu_e \simeq  25.8$ mag arcsec$^{-2}$. Located  2\fdg9 (830~Mpc) from M87 in the SSW direction, and 1\fdg7 (490~Mpc) from M49 in the NNW direction, it has a complex structure with clumpy and twisted isophotes, including possible streams, or ripples, that are suggestive of a post-merger system. A number of GC candidates are associated with the galaxy, and a faint NGVS dwarf (NGVSJ12:27:20.56+09:35:30.1) is located 2\farcm8 (13~kpc) to the W.
    
    \item {\tt NGVSUDG-10.} This galaxy, VCC1052,  was discovered by \citet{Bin85} who not only listed it among their sample of dwarfs of very large size and low surface brightness, but highlighted it as an extreme case within this class (see their Table~XIV). It is located 0\fdg7 (200~kpc) from M87, in the W direction. Although it is a relatively bright galaxy ($g' \simeq 16.1$~mag, $R_e \simeq 47\arcsec$ and $\mu_e \simeq 27.0$ mag arcsec$^{-2}$), no radial velocity measurement exists in the literature. Several other VCC galaxies are found within 13$^\prime$, mainly toward the S and SW: VCC1047 (724 \kms), VCC1036 (1124 \kms) and VCC1010 (934 \kms). Like some other UDGs, it shows a peculiar morphology, including possible spiral arms or tidal streams with triggered star formation, suggestive of a possible past merger. It is a slightly flattened system, with $q\sim0.8$ and $\theta \sim 42^\circ$, and contains a number of GC candidates. A very faint companion from the NGVS (NGVSJ12:27:53.17+12:22:58.8) is located just 1\farcm2 (5.8~kpc) to the NW.
     
    \item {\tt NGVSUDG-11.} This galaxy, VLSB-B, was discovered by \citet{Mih15}. It is a faint system ($g'=18.4$~mag, $R_e=31\arcsec$ and $\mu_e=28.3$ mag arcsec$^{-2}$) located 0\fdg7 (200~kpc) W of M87, near the centroid of the Virgo Cluster galaxy distribution.  A number of GC candidates are associated with the galaxy, although this is a region of high GC density. It is a slightly flattened system, with $q\sim0.8$ and $\theta \sim 118^\circ$. \citet{Tol18} obtained velocities for 4 GCs in this galaxy, reporting a mean velocity of $v_r = 25^{+22}_{-36}$~\kms~ and a velocity dispersion of $\sigma = 47^{+53}_{-29}$~\kms, which potentially implies a very high mass-to-light ratio ($M/L_V\simeq 400$). The nearest VCC galaxy is VCC1104, located 7\farcm2 (35~kpc) to the NE. 
    
    \item {\tt NGVSUDG-12.} This previously undiscovered galaxy is also located in the core region, offset by 0\fdg7 (200~kpc) from M87 in the NW direction. It is a faint system ($g'=18.8$~mag, $R_e=29\arcsec$ and $\mu_e=28.5$ mag arcsec$^{-2}$) with a handful of associated GC candidates. It is also a flattened object, with $q\sim0.6,$ and $\theta\sim117^\circ$. {\tt NGVSUDG-11} is located $\sim$11$^\prime$ (52~kpc) to the SW.
    
    \item {\tt NGVSUDG-13.} Another previously uncatalogued object, this galaxy is located 2\fdg7 (780~Mpc) to the S of M87 and 1\fdg7 (490~Mpc) N of M49. It is of intermediate brightness ($g'=17.1$~mag, $R_e=36\arcsec$ and $\mu_e=27.3$ mag arcsec$^{-2}$) but visibly flattened, with $q\sim0.6,$ and $\theta\sim88^\circ$. At most, there are only one or two associated GC candidates. A possible companion galaxy (NGVSJ12:30:12.55+09:42:56.3) is located very nearby (1$^\prime \sim 5$~kpc), to the E and along the major axis.
    
    \item {\tt NGVSUDG-14.} This is VCC1287 ($v_r = 1071\pm15$ \kms), which was discovered by \citet{Rea83}. It was also catalogued by \citet{Bin85} and highlighted as an extreme case in their sample of large and low surface brightness dwarfs. It is among the most well studied of UDGs, having been targeted by \citet{Bea16}, \citet{Pan18}, and \citet{Gan20}. It is both bright and extended ($g'=15.7$~mag, $R_e=46\arcsec$ and $\mu_e=26.6$ mag arcsec$^{-2}$), and a number of candidate GCs appear to be associated with the galaxy. It is also a very round system, with $q\sim1$ and $\theta\sim20^\circ$. It is located 1\fdg6 (460~kpc) N of M87. A UCD identified by the NGVS (i.e., UCD371 from \citealt{Liu20}) is found nearby, $\sim$106\arcsec (85~kpc) to the WNW.
    
    \item {\tt NGVSUDG-15.} This object was previously identified as VLSB-C by \citet{Mih15}. It is located roughly midway between M87 and M49, with M87  2\fdg0 (580~kpc) to the N and M49 2\fdg4 (690~kpc) to the S. A handful of GC candidates may be associated with the galaxy, including an exceptionally bright candidate ($g'\sim19.5$~mag) located $\sim80\arcsec$ away, in the NW direction. It is a moderately flattened object, with $q\sim0.8$ and $\theta\sim120^\circ$.
    
    \item {\tt NGVSUDG-16.} This previously uncatalogued LSB galaxy is located in the Virgo core, just 0\fdg24 (69~kpc) from M87 in the ESE direction. It is a faint and yet fairly compact system ($g'=19.75$~mag, $R_e=21\arcsec$ and $\mu_e=28.8$ mag arcsec$^{-2}$) with 
 a few GC candidates (although the surface density of GCs is high in this region). It is also a very elongated system, with $q\sim0.3$ and $\theta \sim 72^\circ$.
    
    \item {\tt NGVSUDG-17.} This is another previously uncatalogued LSB galaxy, also located in core region, offset from M87 by just 0\fdg38 (110~kpc) in the ESE direction. A handful of GCs may be associated with the galaxy, although the GC surface density this close to M87 is high. UCD509 from \citet{Liu20} is located just 48$\arcsec$ (38~kpc) to the NE. It is one of the most extreme objects in our sample ($g'=19.2$~mag, $R_e=37\arcsec$ and $\mu_e=29.75$ mag arcsec$^{-2}$), and its axial ratio and orientation are both uncertain.
    
    \item {\tt NGVSUDG-18.} One of the faintest galaxies in our sample, this newly discovered LSB dwarf is located 1\fdg4 (400~kpc) NNE of M87. This is an isolated region of the cluster, with just a handful of Virgo member galaxies nearby. It is an elongated system, with $q\sim0.5$ and $\theta\sim115^\circ$. A couple of GC candidates are found in close proximity.
    
    \item {\tt NGVSUDG-19.} This previously uncatalogued galaxy is an intermediate brightness object, with $g'=17.5$~mag, $R_e=27\arcsec$ and $\mu_e=27.2$ mag arcsec$^{-2}$. It too is located in an isolated region of the cluster, about 2\fdg9 (835~kpc) from M87 in the NNE direction. It is is only slightly flattened, with $q\sim0.9$ and $\theta\sim36^\circ$. About five GC candidates may be associated with the galaxy.
    
    \item {\tt NGVSUDG-20.} This newly discovered galaxy has a large size and low surface brightness ($g'=18.1$~mag, $R_e=43\arcsec$ and $\mu_e=28.8$ mag arcsec$^{-2}$). It is only slightly flattened, with $q\sim0.9$ and $\theta\sim34^\circ$. The galaxy is located in the southern part of the cluster, offset by 1\fdg6 (460~kpc) from M49 in the SE direction. Roughly half dozen of GC candidates appear to be associated with the galaxy.
    
    \item {\tt NGVSUDG-21.} This is another previously uncatalogued LSB galaxy. It is a relatively isolated object, located 2\fdg0 (580~kpc) E of M49. A handful of GC candidates may be associated with the galaxy, which is relatively compact but has a low surface brightness and a faint magnitude ($g'=19.5$~mag, $R_e=19\arcsec$ and $\mu_e=28.4$ mag arcsec$^{-2}$). It is a very round galaxy, with $q \sim 1$.
    
    \item {\tt NGVSUDG-22.} Another previously uncataloged galaxy, this is the faintest object in the {\tt primary} sample ($g'=20.7$~mag, $R_e=16\arcsec$ and $\mu_e=29.2$ mag arcsec$^{-2}$) and close to the NGVS detection limit. Very few GC candidates appear to be associated with the galaxy, whose shape is poorly constrained. It is located in the SE of the cluster, 2\fdg4 (690~kpc) from M87.
    
    \item {\tt NGVSUDG-23.} This is another newly discovered LSB galaxy, found 2\fdg6 (750~kpc) from M49 in the ESE direction. This is a very isolated region of the cluster, with no VCC member galaxies  within 30$^\prime$. There is no obvious GC system associated with the galaxy ($g'=18.1$~mag, $R_e=25\arcsec$ and $\mu_e=27.7$ mag arcsec$^{-2}$). It is an elongated galaxy, with $q\sim0.6$ and $\theta\sim119^\circ$.
    
    \item {\tt NGVSUDG-24.} Another previously uncatalogued galaxy, this object is also found in an isolated part of the cluster, 3\fdg0 (860~kpc) N of M87. It is very faint ($g'=19.5$~mag, $R_e=20\arcsec$ and $\mu_e=28.7$ mag arcsec$^{-2}$) and appears significantly flattened ($q\sim0.7$, $\theta\sim141^\circ$). Another faint NGVS galaxy (NGVSJ12:40:58.86+14:15:57.6) with similar parameters ($g'=19.5$~mag, $R_e=21\arcsec$,  $q\sim0.9$ and $\theta\sim5^\circ$) is visible just 1$\arcmin$ to the NE; together these galaxies may form a LSB pair.
    
    \item {\tt NGVSUDG-25.} This is VCC1884, which was first detected by \citet{Rea83} and noted by \citet{Bin85} as an extreme case of a dwarf of very large size and low surface brightness. It is one of the brighter objects in our sample ($g'=16.2$~mag, $R_e=39\arcsec$ and $\mu_e=26.7$ mag arcsec$^{-2}$) and is located 3\fdg1 (890~kpc) NE of M49. This moderately flattened object ($q\sim0.8$, $\theta\sim67^\circ$) contains a few candidate GCs, but they are not strongly concentrated on the galaxy.
    
    \item {\tt NGVSUDG-26.} Another galaxy discovered by \citet{Rea83}, this is also VCC2045 ($v_r=1309$~\kms) from \citet{Bin85}. It is a bright dwarf located in the extreme eastern edge of the cluster, 4\fdg5 (1.3~Mpc) from M87, in the ESE direction. It is, along with NGVSUDG-09, one of the brightest objects in our sample ($g'=14.6$~mag, $R_e=41\arcsec$ and $\mu_e=25.9$ mag arcsec$^{-2}$) but satisfies our selection criteria as a LSB outlier within its luminosity class. It is a highly flattened galaxy ($q\sim0.3$, $\theta\sim76^\circ$) with a small number of GC candidates and a prominent nucleus at its photocenter.
    
\end{itemize}

\subsection{Secondary Sample: UDGs}
\label{notes:secondary}

\begin{itemize}
    \item {\tt NGVSUDG-A01.} This isolated galaxy --- a new discovery from the NGVS --- is an outlier in the $L$-$R_e$ relation. It is an intermediate luminosity system ($g'=17.8$~mag, $R_e = 25\arcsec$, $\mu_e= 27.3$ mag arcsec$^{-2}$) and moderately flattened ($q\sim0.8$, $\theta\sim143^\circ$). One or two GC candidates are visible, along with a very bright star $\sim$2\farcm4 to the E. It is located in the NW part of the cluster, 4\fdg7 (1.4~Mpc) from M87 and 7\fdg2 (2.1~Mpc) from M49. 
    
    \item {\tt NGVSUDG-A02.} This newly discovered galaxy --- an outlier in both the $L$-$R_e$ and $L$-$\mu_e$ relations --- is one of the faintest objects in our sample, with $g'=19.8$~mag, $R_e = 17\arcsec$ and $\mu_e=28.6$ mag arcsec$^{-2}$. It is a highly elongated galaxy, with $q\sim0.5$ and $\theta\sim87^\circ$. There is no obvious GC system. It is found in an isolated part of the cluster, 2\fdg6 (760~kpc) WSW of M87. 
    
    \item {\tt NGVSUDG-A03.} Another newly discovered galaxy, this is an intermediate luminosity system ($g'=18.3$~mag, $R_e = 22\arcsec$, $\mu_e=27.5$ mag arcsec$^{-2}$) with nearly round isophotes ($q\sim1.0$). It is an outlier in the $L$-$R_e$ diagram and contains a few GC candidates. The galaxy may also contain a bright nuclear star cluster ($g'=20.7$~mag). It is located  2\fdg2 (640~kpc) WSW of M49. 
    
    \item {\tt NGVSUDG-A04.} Originally discovered by \citet{Rea83}, this is VCC615. It has an intermediate luminosity but a large size, making it an outlier in the $L$-$R_e$ diagram ($g'=17.3$~mag,  $R_e =  26\arcsec$, $\mu_e=26.9$ mag arcsec$^{-2}$). It was previously listed by \citet{Bin85} as an example of a dwarf with large size and low surface brightness. A nearly round galaxy ($q\sim1.0$), it contains roughly a dozen or so candidate GCs. It is located 1\fdg9 (550~kpc) W of M87. \citet{Tol18} measured velocities for seven GCs in this galaxy, finding a mean velocity of $v_r=2094^{+15}_{-13}$~\kms~ and velocity dispersion $\sigma =32^{+17}_{-10}$~\kms. 
    
    \item {\tt NGVSUDG-A05.} This faint, newly detected galaxy ($g'=18.6$~mag, $R_e = 22\arcsec$, $\mu_e=27.8$ mag arcsec$^{-2}$) is moderately flattened ($q\sim0.7$, $\theta\sim175^\circ$) and an outlier in the $L$-$R_e$ and $L$-$\langle\mu\rangle_e$ diagrams. One or two GC candidates are loosely concentrated around the galaxy, which is found 2\fdg1 (600~kpc) NW of M87. Another very faint NGVS galaxy (NGVSJ12:23:49.52+13:34:43.9) lies 1\farcm5 (7~kpc) to the S. 
    
    \item {\tt NGVSUDG-A06.} Another discovery from the NGVS, this is among the faintest objects in our sample and close to the NGVS detection limit ($g'=20.7$~mag, $R_e = 14\arcsec$, $\mu_e=28.8$ mag arcsec$^{-2}$). It is an outlier in the $L$-$R_e$ and $L$-$\langle\mu\rangle_e$ diagrams, and appears to be slightly flattened, with $q\sim0.9$ and $\theta\sim152^\circ$. There is no obvious GC system. The galaxy is located 1\fdg9 (550~kpc) NW of M49. 
    
    \item {\tt NGVSUDG-A07.} This is VCC987, a galaxy listed among the sample of large, low surface brightness dwarfs by \citet{Bin85}. An intermediate luminosity galaxy ($g'=18.0$~mag, $R_e = 24\arcsec$, $\mu_e=27.3$ mag arcsec$^{-2}$) with moderately flattened isophotes ($q\sim0.7$, $\theta\sim0^\circ$), it is an outlier in the $L$-$R_e$ diagram. The galaxy is located in the cluster core, just 0\fdg9 (260~kpc) to the WNW of M87 (i.e., close to the centroid of the Virgo dwarf galaxy population). There is a possible tidal connection to NGC4425 (VCC984; $v_r = 1908\pm5$ \kms) which lies 4\farcm4 (21~kpc) to the N. In addition, a bright UCD is located very nearby (UCD234; \citealt{Liu20}), $35\arcsec$ (2.8~kpc) in the NW direction, inside the putative tidal stream that extends to VCC984. A small number of GC candidates may belong to the galaxy, with a few more possibly associated with the stream. 
    
    \item {\tt NGVSUDG-A08.} This is VLSB-A from \citet{Mih15}. In the NGVS images, this very low surface brightness object appears relatively compact ($g'=19.1$~mag, $R_e=15\arcsec$, $\mu_e=28.5$ mag arcsec$^{-2}$) but nevertheless an outlier in the $L$-$\mu_e$ diagram. It contains a prominent nucleus ($g' = 20.6$~mag) with a measured velocity of $-120\pm40$~\kms\ (Ko et al., in preparation). Although it appears to have elongated isophotes ($q\sim0.6$, $\theta\sim105^\circ$), the diffuse outer structure appears complex and may not well represented by a simple S\'ersic model: i.e., \citet{Mih15} report a significantly larger effective radius and brighter magnitude from Burrell Schmidt imaging. We suspect this object may be a UCD caught in the act of formation, with the bright nucleus being stripped from a diffuse and possibly irregular, extended envelope.  A few GC candidates are also visible, although the surface density of GCs is high in this field, which is located just 0\fdg77 (220~kpc) NW of M87. 
    
    \item {\tt NGVSUDG-A09.} Originally discovered by \citet{Rea83}, this is VCC1249 ($276\pm1$ \kms). It is a bright ($g'=14.3$~mag, $R_e=37\arcsec$, $\mu_e=24.8$ mag arcsec$^{-2}$), well studied object (see, e.g., \citealt{Arr12}) that is clearly interacting with M49, which is located just 5\farcm5 (26~kpc) to the NW. It is a flattened system, with $q\sim0.7$ and $\theta\sim179^\circ$, and a highly disturbed morphology due to its strong tidal encounter with M49. It is unusual in being a gas-rich system with embedded HII regions whose formation was triggered by the encounter. Although its surface brightness is not exceptionally low, it is nevertheless  an outlier in the $L$-$R_e$ and $L$-$\langle\mu\rangle_e$ diagrams due its large radius. While not a UDG in the traditional sense, its appearance here illustrates how tidal interactions can dictate the UDG classification. 
    
    \item {\tt NGVSUDG-A10.} This is VCC1448 = IC3475 ($v_r = 2583\pm11$ \kms), an ImIV/dE1$_{\rm p}$ galaxy that was cataloged by \citet{Rea83} and noted by \citet{Bin85} as an extreme example of a large, low surface brightness galaxy. It is, in fact, one of the brightest UDGs in our sample, with $g'=13.7$~mag, $R_e=44\arcsec$ and $\mu_e=24.6$ mag arcsec$^{-2}$. Located  0\fdg6 (170~kpc) NE of M87, with a number of other low-luminosity Virgo galaxies nearby, it is an extraordinary galaxy, unlike most others in our sample. It is flattened ($q\sim0.8$, $\theta\sim69^\circ$), with isophotes that show a number of sharp twists and shells. It also contains a rich system of GC candidates that stands out sharply against the background, as well as a possible nuclear star cluster. This is likely a post-merger, or post-interaction, system. It is an outlier in both the $L$-$R_e$ and $L$-$\langle\mu\rangle_e$ relations. 
    
    \item {\tt NGVSUDG-A11.} This is VCC1681, a bright galaxy with a somewhat large size for its luminosity ($g'=16.4$~mag, $R_e=29\arcsec$, $\mu_e=26.5$ mag arcsec$^{-2}$). An outlier in both the $L$-$R_e$ and $L$-$\mu_e$ diagrams, this is a moderately flattened object, with $q\sim0.8$ and $\theta\sim112^\circ$. It contains one or two GC candidates as well as a prominent central nucleus ($g'\sim21.5$~mag). A faint spiral pattern is visible at large radius. VCC1684 is located 3\farcm5 (17~kpc) to the S, while the very bright interacting pair (VCC1673 and VCC1676) are found 5\farcm8 (28~kpc) to the N. The galaxy is located 1.9 (540~kpc) from M87 in the SE direction.  
        
    \item {\tt NGVSUDG-A12.} A previously undiscovered galaxy, this is the faintest object in our sample and close to the NGVS detection limit ($g'=20.8$~mag, $R_e = 13\arcsec$, $\mu_e=29.1$ mag arcsec$^{-2}$). An outlier in the $L$-$\mu_e$ diagram, the galaxy appears flattened, with $q\sim0.7$ and $\theta\sim73^\circ$, although these parameters should be viewed with caution given its extreme faintness. There is no perceptible GC system. It is located 2\fdg1 (600~kpc) from M87 in the ESE direction. 
         
    \item {\tt NGVSUDG-A13.} This is VCC1835, a faint LSB dwarf ($g'=18.2$~mag, $R_e = 25\arcsec$, $\mu_e=27.6$ mag arcsec$^{-2}$) that is an outlier in the $L$-$R_e$ and $L$-$\langle\mu\rangle_e$ diagrams. It is significantly flattened ($q\sim0.6$ and $\theta\sim88^\circ$). It is a relatively isolated object, offset by 2\fdg3 (660~kpc) from M87 in the ENE direction. Two nearby bright stars may complicate the isophotal analysis. 
    \item {\tt NGVSUDG-A14.} This newly discovered galaxy ($g'=19.5$~mag, $R_e = 20\arcsec$, $\mu_e=28.3$ mag arcsec$^{-2}$) is an outlier in the $L$-$R_e$ and $L$-$\langle\mu\rangle_e$ diagrams. It is slightly flattened ($q\sim0.9$ and $\theta\sim5^\circ$) and appears to have no obvious GC system. Another very faint ($g'=19.5$~mag) NGVS galaxy (NGVSJ12:40:56.41+14:15:16.3) is located $\sim$0\farcm75 (3.6 kpc) to the S, and together these galaxies may form a binary pair. The galaxy is located in a low-density region of the cluster, 3\fdg1 (890~kpc) from M87 in the NE direction. 

    \item {\tt NGVSUDG-A15.} This is VCC1882, a faint LSB galaxy with $g'=18.6$~mag, $R_e = 19\arcsec$ and $\mu_e=27.9$ mag arcsec$^{-2}$. It is an outlier in the $L$-$\mu_e$ diagram and a highly flattened object, with $q\sim0.5$ and $\theta\sim71^\circ$. There are a handful of GC candidates present, as well as a prominent nucleus ($g'=21.7$) at the photocenter. It is located 2\fdg7 (780~kpc) SE of M87 and roughly$\sim8\arcmin$ (38~kpc) to the WNW of M59 (VCC1903). 

    \item {\tt NGVSUDG-A16.} This newly discovered LSB galaxy ($g'=18.3$~mag, $R_e = 23\arcsec$, $\mu_e=27.6$ mag arcsec$^{-2}$) is significantly flattened, with $q\sim0.6$ and $\theta\sim87^\circ$. It is an outlier in the $L$-$R_e$ diagram. There is no obvious GC system. The galaxy is located 3\fdg1 (890~kpc) E of M87, and 5\farcm6 (27~kpc) N of VCC1943, a spiral galaxy that forms a pair with VCC1931. 
    
    \item {\tt NGVSUDG-A17.} Another NGVS detection, this faint galaxy ($g'=19.4$~mag, $R_e = 21\arcsec$, $\mu_e=28.2$ mag arcsec$^{-2}$) is significantly flattened ($q\sim0.7$, $\theta\sim98^\circ$). An outlier in the $L$-$R_e$ and $L$-$\langle\mu\rangle_e$ diagrams, it contains few GC candidates. It is located 5\fdg1 (1.5~Mpc) NE of M87, in a very sparse region of the cluster, although NGC4651 ($v_r=877\pm2$ \kms) is located $\sim10\arcmin$ to the SE. 
   
    \item {\tt NGVSUDG-A18.} An outlier in the $L$-$R_e$ diagram and another NGVS discovery ($g'=18.8$~mag, $R_e = 21\arcsec$, $\mu_e=28.0$ mag arcsec$^{-2}$), this galaxy is significantly flattened ($q\sim0.7$, $\theta\sim67^\circ$) and may contain a small number of GC candidates. It is located 3\fdg5 (1~Mpc) NE of M87, in an isolated region of the cluster. 
    
\end{itemize}

\subsection{Secondary Sample: Late-Type Galaxies}
\label{notes:spirals}

\begin{itemize}
    
    \item {\tt NGVSUDG-B01.} This is NGC4269 (VCC373), a bright S0 galaxy ($v_r = 2074\pm19$ \kms) located close to VCC366, another S0 galaxy. A very bright star is located  1\farcm3 to the north, which further complicates the fitting, although the isophote residuals are well behaved. It is an outlier in the $L$-$R_e$ diagram. 
    
    \item {\tt NGVSUDG-B02.} This is NGC4411 (VCC905), a bright, face-on spiral galaxy of type SB(rs)c ($v_r = 1278\pm1$ \kms). The galaxy has a diameter of $\sim$ 4\farcm5 (22~kpc) and is an outlier in the $L$-$\mu_e$ diagram. There are several other galaxies nearby, including NGC4411b (see below), and VCC933 and VCC902 to the south. 
    
    \item {\tt NGVSUDG-B03.} This is NGC4411b (VCC939), a large, bright and face-on spiral galaxy of type SAB(s)cd;Sy ($v_r = 1272\pm1$ \kms). The extended spiral arms span a diameter of $\sim$6\farcm8 (33 kpc). It is an outlier in the $L$-$R_e$ and $L$-$\langle\mu\rangle_e$ diagrams. 
    
    \item {\tt NGVSUDG-B04.} The is NGC4421 (VCC966), a bright SB(s)0/a galaxy ($v_r = 1603\pm10$ \kms) located 3\fdg2 (920~Mpc) to the NNW of M87. Although it is an isolated object, a very bright star is located 2\farcm7 to the NW, which may affect the isophotal fit. It is an outlier in the $L$-$\mu_e$ diagram. 
    
    \item {\tt NGVSUDG-B05.} This is VCC975 ($v_r = 933\pm4$ \kms), a face-on Sm-type galaxy of low surface brightness. It is located $\sim$1$^{\circ}$ to the SW of M49, with two bright stars within 2\farcm7 that may affect the isohotal fit. It is an outlier in both the $L$-$R_e$ and $L$-$\langle\mu\rangle_e$ diagrams. 
    
    \item {\tt NGVSUDG-B06.} This is NGC4523 (VCC1524), a face-on SAB(s)m galaxy ($v_r = 262\pm4$ \kms) located 1\fdg3 (375~Mpc) NNE of M87. The galaxy spans a diameter of $\sim$5$^\prime$ (24~kpc). It is an outlier in both the $L$-$R_e$ and $L$-$\langle\mu\rangle_e$ diagrams. 
    
    \item {\tt NGVSUDG-B07.} This is NGC4535 (VCC1555), a bright, edge-on SAB(s)c galaxy ($v_r =1964\pm1$ \kms) located $\sim$1\fdg1 (315~Mpc) to the ENE of M49. The galaxy spans a diameter of $\sim$10$^\prime$ (48~kpc). It is an outlier in both the $L$-$R_e$ and $L$-$\langle\mu\rangle_e$ diagrams. 
    
    \item {\tt NGVSUDG-B08.} This is NGC4689 (VCC2058), an edge-on spiral galaxy of type SA(rs)bc, with $v_r = 1616\pm5$ \kms. The galaxy is located in a very isolated region of the cluster, 4\fdg3 (1.2~Mpc) from M87 in the ENE direction. The galaxy spans a diameter of $\sim$8.8$^\prime$ (42~kpc). It is an outlier in both the $L$-$R_e$ and $L$-$\langle\mu\rangle_e$ diagrams. 
        
\end{itemize}

\label{sec:appendix}

\facility{CFHT}

\scriptsize


\clearpage

\tabletypesize{\scriptsize}
\begin{deluxetable*}{lcrcccccl}
\tablecaption{Ultra-Diffuse Galaxies in the Virgo Cluster: Primary Sample\label{tbl:udgcat}}
\tablehead{
\colhead{Name} & \colhead{$\alpha_{\rm J2000}$} & \colhead{$\delta_{\rm J2000}$} & \colhead{$m_{g'}$} & \colhead{$R_{\rm e}$} & \colhead{$\mu_{e}$} & \colhead{$\langle\mu\rangle_{e}$} & \colhead{Other name} & \colhead{NGVS IAU Name} \\
\colhead{} & \colhead{(deg)} & \colhead{(deg)} & \colhead{(mag)} & \colhead{(arcsec)} & \colhead{(mag arcsec$^{-2}$)} & \colhead{(mag arcsec$^{-2}$)} & \colhead{} & \colhead{}
}
\startdata
NGVSUDG-01 & 184.1357917 & 13.1642500 & $16.78$ & $ 62.94$ & $28.89$ & $27.75$ &  VCC197$^\dagger$ & NGVSJ12:16:32.59+13:09:51.3 \\
NGVSUDG-02 & 184.9040000 & 15.4546667 & $16.61$ & $ 33.93$ & $26.80$ & $26.25$ &  VCC360$^\dagger$ & NGVSJ12:19:36.96+15:27:16.8 \\
NGVSUDG-03 & 186.0076667 & 13.8656667 & $18.39$ & $ 32.75$ & $28.32$ & $27.95$ &         & NGVSJ12:24:01.84+13:51:56.4 \\
NGVSUDG-04 & 186.1752500 & 13.5168333 & $17.60$ & $ 32.65$ & $27.85$ & $27.15$ &  VLSB-D$^\star$ & NGVSJ12:24:42.06+13:31:00.6 \\
NGVSUDG-05 & 186.4067073 & 10.2496165 & $16.96$ & $ 33.96$ & $27.10$ & $26.60$ &  VCC811 & NGVSJ12:25:37.61+10:14:58.6 \\
NGVSUDG-06 & 186.6593632 & 13.0789384 & $18.19$ & $ 25.93$ & $28.15$ & $27.25$ &  VCC927 & NGVSJ12:26:38.25+13:04:44.2 \\
NGVSUDG-07 & 186.7015070 & 13.3549078 & $18.97$ & $ 27.86$ & $28.88$ & $28.19$ &  N lsb10$^\diamond$ & NGVSJ12:26:48.36+13:21:17.7 \\
NGVSUDG-08 & 186.8156158 & 13.4489043 & $18.91$ & $ 20.88$ & $28.28$ & $27.51$ &         & NGVSJ12:27:15.75+13:26:56.1 \\
NGVSUDG-09 & 186.8814384 &  9.5956422 & $14.55$ & $ 53.69$ & $25.83$ & $25.19$ & VCC1017$^\dagger$ & NGVSJ12:27:31.55+09:35:44.3 \\
NGVSUDG-10 & 186.9800778 & 12.3692963 & $16.06$ & $ 47.44$ & $27.00$ & $26.43$ & VCC1052 & NGVSJ12:27:55.22+12:22:09.5 \\
NGVSUDG-11 & 187.0419691 & 12.7248463 & $18.42$ & $ 31.27$ & $28.31$ & $27.89$ &  VLSB-B & NGVSJ12:28:10.07+12:43:29.4 \\
NGVSUDG-12 & 187.1578363 & 12.8616713 & $18.75$ & $ 28.63$ & $28.46$ & $28.02$ &         & NGVSJ12:28:37.88+12:51:42.0 \\
NGVSUDG-13 & 187.5361667 &  9.7156111 & $17.09$ & $ 35.76$ & $27.27$ & $26.84$ &         & NGVSJ12:30:08.68+09:42:56.2 \\
NGVSUDG-14 & 187.6017905 & 13.9818128 & $15.71$ & $ 45.84$ & $26.58$ & $26.01$ & VCC1287$^\dagger$ & NGVSJ12:30:24.43+13:58:54.5 \\
NGVSUDG-15 & 187.6554167 & 10.3480556 & $17.45$ & $ 46.54$ & $28.25$ & $27.77$ &  VLSB-C$^\star$ & NGVSJ12:30:37.30+10:20:53.0 \\
NGVSUDG-16 & 187.9500400 & 12.3591964 & $19.75$ & $ 21.22$ & $28.83$ & $28.38$ &         & NGVSJ12:31:48.01+12:21:33.1 \\
NGVSUDG-17 & 188.0938456 & 12.3255700 & $19.20$ & $ 37.24$ & $29.75$ & $29.05$ &         & NGVSJ12:32:22.52+12:19:32.1 \\
NGVSUDG-18 & 188.2591670 & 13.7038890 & $20.03$ & $ 22.17$ & $29.13$ & $28.74$ &         & NGVSJ12:33:02.20+13:42:14.0 \\
NGVSUDG-19 & 188.3732917 & 15.2341111 & $17.51$ & $ 27.20$ & $27.24$ & $26.67$ &         & NGVSJ12:33:29.59+15:14:02.8 \\
NGVSUDG-20 & 188.8035833 &  7.0562222 & $18.09$ & $ 43.48$ & $28.82$ & $28.26$ &         & NGVSJ12:35:12.86+07:03:22.4 \\
NGVSUDG-21 & 189.4572083 &  7.8230833 & $19.50$ & $ 18.65$ & $28.37$ & $27.84$ &         & NGVSJ12:37:49.73+07:49:23.1 \\
NGVSUDG-22 & 189.5413750 & 10.7882222 & $20.66$ & $ 15.64$ & $29.25$ & $28.61$ &         & NGVSJ12:38:09.93+10:47:17.6 \\
NGVSUDG-23 & 189.9500000 &  7.3131111 & $18.12$ & $ 25.49$ & $27.68$ & $27.13$ &         & NGVSJ12:39:48.00+07:18:47.2 \\
NGVSUDG-24 & 190.2350417 & 14.2545278 & $19.46$ & $ 20.30$ & $28.66$ & $27.98$ &         & NGVSJ12:40:56.41+14:15:16.3 \\
NGVSUDG-25 & 190.4138991 &  9.2084609 & $16.20$ & $ 39.28$ & $26.69$ & $26.16$ & VCC1884$^\dagger$ & NGVSJ12:41:39.34+09:12:30.5 \\
NGVSUDG-26 & 191.7311710 & 10.1824199 & $14.61$ & $ 41.16$ & $25.87$ & $24.68$ & VCC2045$^\dagger$ & NGVSJ12:46:55.48+10:10:56.7 \\
\enddata
\tablenotetext{\dagger}{~Galaxy first detected in \citet{Rea83}.}
\tablenotetext{\star}{~Galaxy first detected in \citet{Mih15}.}
\tablenotetext{\diamond}{~Galaxy first detected in \citet{Cal06}.}
\end{deluxetable*}
\clearpage

\begin{deluxetable*}{lcccccc}
\tablecaption{Globular Cluster Properties of Ultra-Diffuse Galaxies\label{tbl:gcsn}}
\tablehead{
\colhead{Name} & \colhead{$M_V$} & \colhead{$N_{GC,raw}$} & \colhead{$N_{GC,corr}$} & \colhead{$S_N$} & \colhead{Nucleus} & \colhead{Notes} \\
\colhead{} & \colhead{[mag]} & \colhead{ } & \colhead{ } & \colhead{ } & \colhead{ } & \colhead{} \\
\colhead{(1)} & \colhead{(2)} & \colhead{(3)} & \colhead{(4)} & \colhead{(5)} & \colhead{(6)} & \colhead{(7)}
}
\startdata
 NGVSUDG-01  & $-14.5$ & $  4$ & $  -8.8 \pm    6.2$ & $ -13.9 \pm    9.7$ & yes & \nodata \\
 NGVSUDG-02  & $-14.7$ & $  2$ & $   1.1 \pm    4.3$ & $   1.5 \pm    5.9$ & no  & \nodata \\
 NGVSUDG-03  & $-12.9$ & $  3$ & $   4.4 \pm    5.0$ & $  30.9 \pm   34.7$ & no  & \nodata \\
 NGVSUDG-04  & $-13.7$ & $  6$ & $  13.0 \pm    6.9$ & $  43.7 \pm   23.2$ & yes & \nodata \\
 NGVSUDG-05  & $-14.3$ & $  8$ & $  15.8 \pm    8.4$ & $  29.7 \pm   15.8$ & no  & \nodata \\
 NGVSUDG-06  & $-13.1$ & $  5$ & $   7.3 \pm    7.4$ & $  42.4 \pm   43.0$ & yes & \nodata \\
 NGVSUDG-07  & $-12.3$ & $  1$ & $  -0.5 \pm    3.0$ & $  -6.2 \pm   35.6$ & no  & \nodata \\
 NGVSUDG-08  & $-12.4$ & $  1$ & $  -0.9 \pm    4.2$ & $ -10.3 \pm   47.8$ & no  & 2$R_{e,gal}$ aperture is used \\
 NGVSUDG-09  & $-16.7$ & $ 13$ & $  16.5 \pm   11.2$ & $   3.3 \pm    2.3$ & no  & \nodata \\
 NGVSUDG-10  & $-15.2$ & $ 12$ & $  17.9 \pm   11.5$ & $  14.5 \pm    9.4$ & no  & \nodata \\
 NGVSUDG-11  & $-12.9$ & $ 12$ & $  26.1 \pm    9.9$ & $ 187.1 \pm   71.1$ & no  & \nodata \\
 NGVSUDG-12  & $-12.5$ & $  4$ & $   4.5 \pm    6.1$ & $  43.3 \pm   59.5$ & no  & \nodata \\
 NGVSUDG-13  & $-14.2$ & $  1$ & $  -2.4 \pm    3.7$ & $  -5.1 \pm    7.8$ & no  & \nodata \\
 NGVSUDG-14  & $-15.6$ & $ 14$ & $  27.6 \pm   11.1$ & $  16.3 \pm    6.6$ & no  & \nodata \\
 NGVSUDG-15  & $-13.8$ & $  3$ & $  -3.2 \pm    7.6$ & $  -9.3 \pm   22.5$ & no  & \nodata \\
 NGVSUDG-16  & $-11.5$ & $  3$ & $   0.2 \pm    7.9$ & $   4.5 \pm  192.5$ & no  & \nodata \\
 NGVSUDG-17  & $-12.1$ & $  4$ & $  -0.4 \pm    8.1$ & $  -6.5 \pm  120.2$ & no  & \nodata \\
 NGVSUDG-18  & $-11.3$ & $  2$ & $   3.3 \pm    4.0$ & $ 104.6 \pm  126.4$ & no  & \nodata \\
 NGVSUDG-19  & $-13.8$ & $  7$ & $  16.8 \pm    7.5$ & $  52.2 \pm   23.2$ & no  & \nodata \\
 NGVSUDG-20  & $-13.2$ & $  7$ & $  11.3 \pm    8.6$ & $  59.8 \pm   45.4$ & no  & \nodata \\
 NGVSUDG-21  & $-11.8$ & $  3$ & $   6.5 \pm    4.9$ & $ 126.8 \pm   95.0$ & no  & \nodata \\
 NGVSUDG-22  & $-10.6$ & $  1$ & $   1.4 \pm    2.8$ & $  76.4 \pm  157.9$ & no  & \nodata \\
 NGVSUDG-23  & $-13.2$ & $  3$ & $   5.0 \pm    5.0$ & $  27.4 \pm   27.1$ & no  & \nodata \\
 NGVSUDG-24  & $-11.8$ & $  2$ & $   4.0 \pm    4.0$ & $  75.5 \pm   74.3$ & no  & \nodata \\
 NGVSUDG-25  & $-15.1$ & $  2$ & $  -1.9 \pm    5.1$ & $  -1.8 \pm    4.7$ & no  & \nodata \\
 NGVSUDG-26  & $-16.7$ & $  7$ & $  13.4 \pm    8.0$ & $   2.9 \pm    1.7$ & yes & \nodata \\
\enddata
\tablenotetext{}{(1) Name of UDGs}
\tablenotetext{}{(2) $V$-band absolute magnitudes derived from $g$-band magnitudes with assumed constants $(g-V)$ color of 0.1 mag and distance modulus of $31.1$.}
\tablenotetext{}{(3) $N_{GC,raw}$ is the observed number of GC candidates within $1.5R_e$ except for NGVSUDG-08.}
\tablenotetext{}{(4) $N_{GC,corr}$ is total number of GCs after correction for background contamination, areal coverage, and limiting magnitude.}
\tablenotetext{}{(5) Specific frequency estimated with values in column 2 and 4}
\tablenotetext{}{(6) Candidate of nucleated UDG}
\tablenotetext{}{(7) Notes for special conditions} 
\end{deluxetable*}

\begin{deluxetable*}{lcrcccccl}
\tablecaption{Additional Ultra-Diffuse Galaxies in the Virgo Cluster: Secondary Sample\label{tbl:udgcata}}
\tablehead{
\colhead{Name} & \colhead{$\alpha_{\rm J2000}$} & \colhead{$\delta_{\rm J2000}$} & \colhead{$m_{g'}$} & \colhead{$R_{\rm e}$} & \colhead{$\mu_{e}$} & \colhead{$\langle\mu\rangle_{e}$} & \colhead{Other name} & \colhead{NGVS IAU Name} \\
\colhead{} & \colhead{(deg)} & \colhead{(deg)} & \colhead{(mag)} & \colhead{(arcsec)} & \colhead{(mag arcsec$^{-2}$)} & \colhead{(mag arcsec$^{-2}$)} & \colhead{} & \colhead{}
}
\startdata
\hline
\multicolumn{9}{c}{{\it Ultra-Diffuse Galaxies}} \\ 
\hline
NGVSUDG-A01 & 183.0884684 & 13.7379406 & $17.76$ & $ 25.23$ & $27.30$ & $26.75$ &         & NGVSJ12:12:21.23+13:44:16.6 \\
NGVSUDG-A02 & 185.0474167 & 11.8990278 & $19.78$ & $ 17.43$ & $28.65$ & $27.96$ &         & NGVSJ12:20:11.38+11:53:56.5 \\
NGVSUDG-A03 & 185.5149167 & 11.7215278 & $18.34$ & $ 22.44$ & $27.50$ & $27.08$ &         & NGVSJ12:22:03.58+11:43:17.5 \\
NGVSUDG-A04 & 185.7691250 & 12.0148333 & $17.25$ & $ 26.33$ & $26.86$ & $26.34$ &  VCC615$^\dagger$ & NGVSJ12:23:04.59+12:00:53.4 \\
NGVSUDG-A05 & 185.9472080 & 13.6023060 & $18.61$ & $ 21.72$ & $27.79$ & $27.28$ &         & NGVSJ12:23:47.33+13:36:08.3 \\
NGVSUDG-A06 & 186.6558750 &  9.7422222 & $20.70$ & $ 14.35$ & $28.80$ & $28.47$ &         & NGVSJ12:26:37.41+09:44:32.0 \\
NGVSUDG-A07 & 186.8144228 & 12.6615036 & $18.04$ & $ 23.74$ & $27.27$ & $26.91$ &  VCC987 & NGVSJ12:27:15.46+12:39:41.4 \\
NGVSUDG-A08 & 187.0660464 & 12.8700124 & $19.09$ & $ 14.93$ & $28.53$ & $26.96$ &  VLSB-A$^\star$ & NGVSJ12:28:15.85+12:52:12.0 \\
NGVSUDG-A09 & 187.5025342 &  7.9293817 & $14.25$ & $ 36.76$ & $24.84$ & $24.07$ & VCC1249$^\dagger$ & NGVSJ12:30:00.61+07:55:45.8 \\
NGVSUDG-A10 & 188.1699846 & 12.7711479 & $13.67$ & $ 43.92$ & $24.57$ & $23.88$ & VCC1448$^\ddagger$ & NGVSJ12:32:40.80+12:46:16.1 \\
NGVSUDG-A11 & 189.1558388 & 11.1536465 & $16.40$ & $ 28.91$ & $26.53$ & $25.68$ & VCC1681 & NGVSJ12:36:37.40+11:09:13.1 \\
NGVSUDG-A12 & 189.8416250 & 12.0929440 & $20.76$ & $ 12.58$ & $29.10$ & $28.24$ &         & NGVSJ12:39:21.99+12:05:34.6 \\
NGVSUDG-A13 & 190.0883727 & 12.7174544 & $18.23$ & $ 24.53$ & $27.62$ & $27.16$ & VCC1835 & NGVSJ12:40:21.21+12:43:02.8 \\
NGVSUDG-A14 & 190.2452500 & 14.2660000 & $19.50$ & $ 19.53$ & $28.26$ & $27.94$ &         & NGVSJ12:40:58.86+14:15:57.6 \\
NGVSUDG-A15 & 190.3785919 & 11.6821610 & $18.58$ & $ 18.77$ & $27.90$ & $26.93$ & VCC1882 & NGVSJ12:41:30.86+11:40:55.8 \\
NGVSUDG-A16 & 190.7358000 & 13.3470670 & $18.29$ & $ 23.27$ & $27.55$ & $27.11$ &         & NGVSJ12:42:56.59+13:20:49.4 \\
NGVSUDG-A17 & 190.7795420 & 16.4783330 & $19.38$ & $ 21.04$ & $28.24$ & $27.97$ &         & NGVSJ12:43:07.09+16:28:42.0 \\
NGVSUDG-A18 & 190.8370000 & 14.0341389 & $18.77$ & $ 20.84$ & $27.96$ & $27.35$ &         & NGVSJ12:43:20.88+14:02:02.9 \\
\hline
\multicolumn{9}{c}{{\it Late-Type Galaxies}} \\ 
\hline
NGVSUDG-B01 & 184.9549824 &  6.0149443 & $12.53$ & $ 32.74$ & $23.91$ & $22.10$ &  VCC373 & NGVSJ12:19:49.20+06:00:53.8 \\
NGVSUDG-B02 & 186.6245659 &  8.8722006 & $13.06$ & $ 40.01$ & $24.03$ & $23.07$ &  VCC905 & NGVSJ12:26:29.90+08:52:19.9 \\
NGVSUDG-B03 & 186.6967860 &  8.8845984 & $12.73$ & $ 44.38$ & $23.89$ & $22.96$ &  VCC939 & NGVSJ12:26:47.23+08:53:04.6 \\
NGVSUDG-B04 & 186.7605815 & 15.4614832 & $11.46$ & $ 56.67$ & $23.69$ & $22.22$ &  VCC966 & NGVSJ12:27:02.54+15:27:41.3 \\
NGVSUDG-B05 & 186.7970619 &  7.2630312 & $13.12$ & $ 44.44$ & $23.88$ & $23.36$ &  VCC975 & NGVSJ12:27:11.29+07:15:46.9 \\
NGVSUDG-B06 & 188.4497867 & 15.1682539 & $13.13$ & $ 41.58$ & $24.05$ & $23.22$ & VCC1524 & NGVSJ12:33:47.95+15:10:05.7 \\
NGVSUDG-B07 & 188.5846259 &  8.1979321 & $10.54$ & $ 88.94$ & $23.13$ & $22.28$ & VCC1555 & NGVSJ12:34:20.31+08:11:52.6 \\
NGVSUDG-B08 & 191.9399170 & 13.7628365 & $11.10$ & $ 68.32$ & $23.37$ & $22.27$ & VCC2058 & NGVSJ12:47:45.58+13:45:46.2 \\
\enddata
\tablenotetext{\dagger}{~Galaxy first detected in \citet{Rea83}.}
\tablenotetext{\star}{~Galaxy first detected in \citet{Mih15}.}
\tablenotetext{\ddagger}{~Galaxy also detected in \citet{Rea83}.}
\end{deluxetable*}

\begin{deluxetable*}{lcccccc}
\tablecaption{Globular Cluster Properties of Ultra-Diffuse Galaxies in secondary sample\label{tbl:gcsn2}}
\tablehead{
\colhead{Name} & \colhead{$M_V$} & \colhead{$N_{GC,raw}$} & \colhead{$N_{GC,corr}$} & \colhead{$S_N$} & \colhead{Nucleus} & \colhead{Notes} \\
\colhead{} & \colhead{[mag]} & \colhead{ } & \colhead{ } & \colhead{ } & \colhead{ } & \colhead{}\\
\colhead{(1)} & \colhead{(2)} & \colhead{(3)} & \colhead{(4)} & \colhead{(5)} & \colhead{(6)} & \colhead{(7)}
}
\startdata
NGVSUDG-A01  & $-13.5$ & $  2$ & $   2.8 \pm    4.0$ & $  10.8 \pm   15.8$ & no  & \nodata  \\
NGVSUDG-A02  & $-11.5$ & $  1$ & $  -1.2 \pm    1.8$ & $ -23.1 \pm   44.0$ & no  & 3$R_{e,gal}$ aperture is used  \\
NGVSUDG-A03  & $-12.9$ & $  2$ & $   3.0 \pm    4.0$ & $  19.8 \pm   27.0$ & yes & \nodata \\
NGVSUDG-A04  & $-14.0$ & $ 12$ & $  30.3 \pm    9.6$ & $  74.3 \pm   23.6$ & no  & \nodata \\
NGVSUDG-A05  & $-12.7$ & $  1$ & $   0.5 \pm    3.0$ & $   4.6 \pm   25.9$ & no  & \nodata \\
NGVSUDG-A06  & $-10.6$ & $  2$ & $   0.5 \pm    3.3$ & $  28.2 \pm  194.3$ & no  & 2$R_{e,gal}$ aperture is used  \\
NGVSUDG-A07  & $-13.2$ & $  3$ & $   4.7 \pm    5.0$ & $  23.7 \pm   25.3$ & no  & \nodata   \\
NGVSUDG-A08  & $-12.2$ & $  2$ & $   4.0 \pm    4.1$ & $  52.8 \pm   53.9$ & yes & \nodata  \\
NGVSUDG-A09  & $-17.0$ & $ 23$ & $  12.0 \pm   31.9$ & $   1.8 \pm    4.9$ & no  & Highly uncertain due to close distance to M49 \\
NGVSUDG-A10  & $-17.6$ & $ 39$ & $  99.3 \pm   17.6$ & $   8.9 \pm    1.6$ & yes & \nodata \\
NGVSUDG-A11  & $-14.9$ & $  2$ & $   2.6 \pm    4.2$ & $   2.9 \pm    4.7$ & yes & \nodata \\
NGVSUDG-A12  & $-10.5$ & $  2$ & $  -1.6 \pm    0.7$ & $ -98.1 \pm   46.3$ & no  &5$R_{e,gal}$ aperture is used  \\
NGVSUDG-A13  & $-13.0$ & $  2$ & $   2.8 \pm    4.0$ & $  16.9 \pm   23.9$ & no  & \nodata \\
NGVSUDG-A14  & $-11.8$ & $  1$ & $  -0.6 \pm    3.6$ & $ -11.9 \pm   69.7$ & no & 2$R_{e,gal}$ aperture is used   \\
NGVSUDG-A15  & $-12.7$ & $  3$ & $   6.3 \pm    4.9$ & $  52.2 \pm   40.6$ & yes & \nodata \\
NGVSUDG-A16$^a$  & $-13.0$ & $  1$ & $   0.3 \pm    5.3$ & $   2.2 \pm   33.7$ & no & 2$R_{e,gal}$ aperture is used   \\
NGVSUDG-A17$^c$  & $-11.9$ & $  2$ & $  -5.0 \pm    3.5$ & $ -86.2 \pm   60.4$ & no & 5$R_{e,gal}$ aperture is used   \\
NGVSUDG-A18  & $-12.5$ & $  3$ & $   6.5 \pm    4.9$ & $  64.3 \pm   48.8$ & no  & \nodata \\
\enddata
\tablenotetext{}{(1) Name of UDGs}
\tablenotetext{}{(2) $V$-band absolute magnitudes derived from $g$-band magnitudes with assumed constants $(g-V)$ color of 0.1 mag and distance modulus of $31.1$.}
\tablenotetext{}{(3) $N_{GC,raw}$ is the observed number of GC candidates within $1.5R_e$ except for UDGs with speical notes.}
\tablenotetext{}{(4) $N_{GC,corr}$ is total number of GCs after correction for background contamination, areal coverage, and limiting magnitude.}
\tablenotetext{}{(5) Specific frequency estimated with values in column 2 and 4}
\tablenotetext{}{(6) Candidate of nucleated UDG}
\tablenotetext{}{(7) Notes for special conditions}
\end{deluxetable*}

\clearpage

\end{document}